\documentclass[a4paper,UKenglish,cleveref, autoref, thm-restate]{lipics-v2021}
\usepackage{todonotes}
\usepackage{tikz}
\usepackage{verbatim}
\usepackage{environ}
\usepackage[noline,noend,ruled]{algorithm2e}
\usetikzlibrary{snakes}

\makeatletter
\newsavebox{\measure@tikzpicture}
\NewEnviron{scaletikzpicturetowidth}[1]{%
  \def\tikz@width{#1}%
  \def\tikzscale{1}\begin{lrbox}{\measure@tikzpicture}%
  \BODY
  \end{lrbox}%
  \pgfmathparse{#1/\wd\measure@tikzpicture}%
  \edef\tikzscale{\pgfmathresult}%
  \BODY
}
\makeatother

\title{Hardness of Detecting Abelian and Additive Square Factors in Strings} %TODO Please add

\newcommand\twocol[2]{%
\begin{center}%
\begin{minipage}[t]{0.5\textwidth}%
{#1}%
\end{minipage}\hfill%
\begin{minipage}[t]{0.5\textwidth}%
{#2}%
\end{minipage}%
\end{center}}

\newtheorem{fact}{Fact}
\newcommand{\Oh}{\mathcal{O}}

\DeclareMathOperator{\ThreeSUM}{\textsc{3SUM}\xspace}
\DeclareMathOperator{\ConvThreeSUM}{\textsc{Conv3SUM}\xspace}
\DeclareMathOperator{\MidCond}{\mathbf{\mathbf{\mathbf{\Lambda}}}\xspace} % Nie chcę mathbf{H} bo to sugeruje stringa, jak w A, B, S.
\DeclareMathOperator{\MidSUMone}{\textsc{Odd\mbox{-}3DAP}\xspace}
\DeclareMathOperator{\MidSUMthree}{\textsc{3DAP}\xspace}
\DeclareMathOperator{\OddConv}{\textsc{OddConv3SUM}\xspace}
\newcommand{\floor}[1]{\left\lfloor #1 \right\rfloor}

\newcommand{\zero}{\mathtt{0}}

\newcommand{\one}{\mathtt{1}}
\newcommand{\two}{\mathtt{2}}
\newcommand{\three}{\mathtt{3}}
\newcommand{\four}{\mathtt{4}}
\newcommand{\five}{\mathtt{5}}
\newcommand{\six}{\mathtt{6}}

\renewcommand{\aa}{\mathtt{a}}
\newcommand{\bb}{\mathtt{b}}

\newcommand{\xx}{\bullet}

\newcommand{\yy}{\star}

\newcommand{\zz}{\Diamond}
\newcommand{\zez}{\circ}
\renewcommand{\SS}{\mathbf{S}}
\renewcommand{\AA}{\mathbf{A}}
\newcommand{\BB}{\mathbf{B}}

\newcommand{\Parikh}{\mathit{Parikh}}

\nolinenumbers

\newcommand{\Alphabet}{\mathit{Alph}}
 \newcommand{\defproblem}[3]{
  \vspace{2mm}
\noindent\fbox{
  \begin{minipage}{0.96\textwidth}
  #1\\
  \textbf{{Input:}} #2  \\
  \textbf{{Output:}} #3
  \end{minipage}
  }
  \vspace{2mm}
}

\author{Jakub Radoszewski}{University of Warsaw, Poland\and Samsung R\&D Poland}{jrad@mimuw.edu.pl}{https://orcid.org/0000-0002-0067-6401}{}
\author{Wojciech Rytter}{University of Warsaw, Poland}{rytter@mimuw.edu.pl}{https://orcid.org/0000-0002-9162-6724}{}
\author{Juliusz Straszy{\'n}ski} {University of Warsaw, Poland}{jks@mimuw.edu.pl}{0000-0003-2207-0053}{}
\author{Tomasz Wale\'n}{University of Warsaw, Poland}{walen@mimuw.edu.pl}{https://orcid.org/0000-0002-7369-3309}{}
\author{Wiktor Zuba}{University of Warsaw, Poland}{w.zuba@mimuw.edu.pl}{https://orcid.org/0000-0002-1988-3507}{}
\acknowledgements{The authors warmly thank Paweł Gawrychowski and Tomasz Kociumaka for helpful discussions.}

\authorrunning{Jakub Radoszewski, Wojciech Rytter, Juliusz Straszy\'nski, Tomasz Wale\'n, Wiktor Zuba} %TODO mandatory. First: Use abbreviated first/middle names. Second (only in severe cases): Use first author plus 'et al.'
\titlerunning{Hardness of Detecting Abelian and Additive Squares}

\Copyright{Jakub Radoszewski, Wojciech Rytter, Juliusz Straszy\'nski, Tomasz Wale\'n, Wiktor Zuba} %TODO mandatory, please use full first names. LIPIcs license is "CC-BY";  http://creativecommons.org/licenses/by/3.0/


 %TODO mandatory; please add comma-separated list of keywords

\hideLIPIcs

\begin{document}

\maketitle

%TODO mandatory: add short abstract of the document
\begin{abstract}
We prove 3SUM-hardness (no strongly subquadratic-time algorithm, assuming the 3SUM conjecture) of several problems related to finding Abelian square and additive square factors in a string. In particular, we conclude conditional optimality of the state-of-the-art algorithms for finding such factors. 

Overall, we show 3SUM-hardness of (a) detecting an Abelian square factor of an odd half-length, (b) computing centers of all Abelian square factors, (c) detecting an additive square factor in a length-$n$ string of integers of magnitude $n^{\Oh(1)}$, and (d) a problem of computing a double 3-term arithmetic progression (i.e., finding indices $i \ne j$ such that $(x_i+x_j)/2=x_{(i+j)/2}$) in a sequence of integers $x_1,\dots,x_n$ of magnitude $n^{\Oh(1)}$.

Problem (d) is essentially a convolution version of the AVERAGE problem that was proposed in a manuscript of Erickson. We obtain a conditional lower bound for it with the aid of techniques recently developed by Dudek et al.\ [STOC 2020]. Problem (d) immediately reduces to problem (c) and is a step in reductions to problems (a) and (b). In conditional lower bounds for problems (a) and (b) we apply an encoding of Amir et al.\ [ICALP 2014] and extend it using several string gadgets that include arbitrarily long Abelian-square-free strings.

Our reductions also imply conditional lower bounds for detecting Abelian squares in strings over a constant-sized alphabet. We also show a subquadratic upper bound in this case, applying a result of Chan and Lewenstein [STOC 2015].

\end{abstract}

%\pagenumbering{arabic} 
\section{Introduction}
\subparagraph*{Abelian squares.}
An Abelian square, Ab-square in short (also known as a jumbled square), is a string of the form $XY$, where $Y$ is a permutation of $X$; we say that $X$ and $Y$ are Ab-equivalent. We are interested in factors (i.e., substrings composed of consecutive letters) of a given text string being Ab-squares.
\begin{example}\label{ex1} The string

\begin{center}
    \begin{tikzpicture}[scale=0.3]
        \foreach \x/\v in {0/0, 1/6, 2/1, 3/0, 4/5, 5/6, 6/5, 7/1, 8/6, 9/1, 10/0,
11/1, 12/1,13/1,14/1,15/0,16/6,17/5, 18/6, 19/5,20/7,21/8, 22/6, 23/1, 24/6, 25/5, 26/0, 27/5, 
28/1, 29/0, 30/5, 31/6, 32/6, 33/5, 34/0, 35/6, 36/0, 37/3, 38/0, 39/6, 40/5, 41/2}{
            \draw (\x,0) node[above] {\v};
        }
        \begin{scope}[xshift=-0.5cm,yshift=0.7cm]
        \draw[brown,thick] (6,0.7) -- (6,1) -- (18,1) -- (18,0.7)  (12,0.7) -- (12,1);
        \draw[brown,thick,xshift=16cm] (6,0.7) -- (6,1) -- (18,1) -- (18,0.7)  (12,0.7) -- (12,1);
        \end{scope}
    \end{tikzpicture}
\end{center}

\noindent has 
exactly two Ab-square factors of length 12, shown above
(but it has also  Ab-squares of other lengths, 
e.g. 5665, 11, 1111, 011110).
\end{example}

Ab-squares were first studied by Erd\H{o}s \cite{Erdos}, who posed a question on the smallest alphabet size for which there exists an infinite Ab-square-free string, i.e., an infinite string without Ab-square factors. The first example of such a string over a finite alphabet was given by Evdokimov~\cite{evdokimov}. Later the alphabet size was improved to five by Pleasants~\cite{Pleasants} and finally an optimal example over a four-letter alphabet was shown by Ker\"anen~\cite{DBLP:conf/icalp/Keranen92}. Further results on combinatorics of Ab-square-free strings and several examples of their applications in group theory, algorithmic music and cryptography can be found in~\cite{DBLP:journals/tcs/Keranen09} and references therein. Avoidability of long Ab-squares was also considered~\cite{DBLP:journals/siamdm/RaoR18}.

Strings containing Ab-squares were also studied. Motivated by another problem posed by Erd\H{o}s \cite{Erdos}, Entringer et al.~\cite{ENTRINGER1974159} showed that every infinite binary string has arbitrarily long Ab-square factors. Fici et al.~\cite{DBLP:journals/tcs/FiciMS17} considered infinite strings containing many distinct Ab-squares. A string of length $n$ may contain $\Theta(n^2)$ Ab-square factors that are distinct as strings, but contains only $\Oh(n^{11/6})$ Ab-squares which are 
pairwise Abelian nonequivalent (correspond to different Parikh vectors), see~\cite{DBLP:journals/tcs/KociumakaRRW16}. It is also conjectured that a binary string of length $n$ must have at least $\floor{n/4}$ distinct~\cite{crochemore_et_al:DR:2014:4552} and nonequivalent~\cite{DBLP:conf/cpm/FraenkelSP97} Ab-square factors. For more conjectures related to combinatorics of Ab-square factors of strings ad circular strings, see~\cite{simpson2018solved}.

Several algorithms computing Ab-square factors of a string are known. All Ab-squares in a string of length $n$ can be computed in $\Oh(n^2)$ time~\cite{Cummings_weakrepetitions}.
For a string over a constant-sized alphabet, all Ab-square factors of a string can be computed in $\Oh(n^2/\log^2 n+\mathsf{output})$ time and the longest Ab-square can be computed in $\Oh(n^2/\log^2 n)$ time~\cite{DBLP:conf/macis/KociumakaRW15,AIMS}. Moreover, for a string of length $n$ that is given by its run-length encoding consisting of $r$ runs, the longest Ab-square that ends at each position can be computed in $\Oh(|\Sigma| (r^2+n))$ time~\cite{DBLP:journals/tcs/AmirAHLLR16} or in $\Oh(rn)$ time~\cite{DBLP:conf/iwoca/SugimotoNIBT17}; both approaches require $\Omega(n^2)$ time in the worst case.

In \cite{DBLP:journals/combinatorics/RichmondS09} a different problem of enumerating strings being Ab-squares was considered.

\subparagraph*{Additive squares.}
An additive square is an even-length string over an integer alphabet such that the sums of characters of the halves of this string are the same. 
\begin{example} The  following string
has exactly 4 additive squares of length 10, as shown.
\begin{center}
    \begin{tikzpicture}[scale=0.3]
        \foreach \x/\v in {0/1,1/2,2/0,3/3,4/2,5/1,6/2,7/0,8/2,9/3,10/2,11/1,12/0,13/1,14/2,15/3}{
            \draw (\x,0) node[above] {\v};
        }
        \foreach \y/\s in {0/0,0.5/2,1/3,1.5/4}{
            \begin{scope}[xshift=-0.5cm+\s cm,yshift=0.7cm+\y cm]
                \draw[brown,thick] (0,0.7) -- (0,1) -- (10,1) -- (10,0.7)  (5,0.7) -- (5,1);
            \end{scope}
        }
    \end{tikzpicture}
\end{center}
\noindent 
All of them except for the rightmost one are also Ab-squares. This string does not contain any longer additive square. Altogether this string has 8 additive square factors.
\end{example}

An Ab-square (over an integer alphabet) is an additive square, but not necessarily the other way around.
Combinatorially, problems related to additive squares are hard,
in particular avoiding additive squares seems more difficult than 
avoiding Ab-squares. There are infinitely many strings over $\{0,1,2,3\}$
avoiding Ab-squares, but there are only finitely many strings 
over the same alphabet
avoiding additive squares; see \cite{DBLP:journals/int/FreedmanB16}.

In fact
it is unknown if there are infinitely many strings over any finite 
integer alphabet avoiding additive squares~\cite{BF1987,HH2000,PV1998}. For additive cubes the property was proved in~\cite{DBLP:journals/jacm/CassaigneCSS14} (see also~\cite{DBLP:conf/dlt/LietardR20}) however.

Nowadays, combinatorial study of Ab-square and additive square factors often involves computer experiments; see e.g.\ \cite{DBLP:journals/jacm/CassaigneCSS14,DBLP:journals/tcs/FiciMS17,DBLP:journals/siamdm/RaoR18}. In addition to other applications, efficient algorithms detecting such types of squares could provide a significant aid in this research. In case of classic square factors (i.e., factors of the form $XX$), a linear-time algorithm for computing them is known for a string over a constant~\cite{DBLP:journals/jcss/GusfieldS04} and over an integer alphabet~\cite{DBLP:conf/cpm/BannaiIK17,DBLP:journals/tcs/CrochemoreIKRRW14}. We show that, unfortunately, in many cases the existence of near-linear-time algorithms for detecting Ab-square and additive square factors is unlikely, based on conjectured hardness of the $\ThreeSUM$ problem.

\subparagraph*{$\ThreeSUM$ problem.} 
The problem asks if there are distinct elements $a,b,c \in S$ such that $a+b=c$ for a given set $S$ of $n$ integers
; see \cite{DBLP:conf/stoc/Patrascu10}.
It is a general belief that the following conjecture is true for the word-RAM model.

\subparagraph*{$\ThreeSUM$ conjecture:}

There is no $\Oh(n^{2-\epsilon})$ time algorithm for the $\ThreeSUM$ problem, for any constant $\epsilon>0$.

\smallskip
\noindent
A problem with input of size $n$ is called $\ThreeSUM$-hard if an $\Oh(n^{2-\varepsilon})$-time solution to the problem implies an $\Oh(n^{2-\varepsilon'})$-time solution for $\ThreeSUM$, for some constants $\varepsilon,\varepsilon'>0$.

\subparagraph*{Our results.}
\begin{itemize}
\item
We show that the problems of computing all centers of Ab-square factors and detecting an odd half-length Ab-square factor, called an \emph{odd Ab-square} (consequently also computing all lengths of Ab-square factors), for a length-$n$ string over an alphabet of size $\omega(1)$, cannot be solved in $\Oh(n^{2-\varepsilon})$ time, for constant $\varepsilon>0$, unless the 3SUM conjecture fails. Weaker conditional lower bounds are also stated in the case of a constant-sized alphabet.
\item
For constant-sized alphabets, we show strongly sub-quadratic algorithms for these problems based on an involved result of \cite{DBLP:conf/stoc/ChanL15} related to jumbled indexing.
\item
En route we prove that detection of a double 3-term arithmetic progression (see~\cite{DBLP:journals/int/Brown14}) and additive squares in a length-$n$ sequence of integers of magnitude $n^{\Oh(1)}$ is $\ThreeSUM$-hard.
\end{itemize}
We obtain deterministic conditional lower bounds from a convolution version of $\ThreeSUM$ that is well-known to be $\ThreeSUM$-hard.

\subparagraph*{Related work.}
In the jumbled indexing problem, we are given a text $T$ and are to answer queries for a pattern specified by a Parikh vector which gives, for each letter of the alphabet, the number of occurrences of this letter in the pattern. For each query, we are to check if there is a factor of the text that is Ab-equivalent to the pattern (existence query) or report all such factors (reporting query). Chan and Lewenstein~\cite{DBLP:conf/stoc/ChanL15} showed a data structure that can be constructed in truly subquadratic expected time and answers existence queries in truly sublinear time for a constant-sized alphabet (deterministic constructions for very small alphabets were also shown). Amir et al.~\cite{DBLP:conf/icalp/AmirCLL14} showed under a 3SUM-hardness assumption that jumbled indexing with existence queries requires $\Omega(n^{2-\varepsilon})$ preprocessing time or $\Omega(n^{1-\delta})$ queries for any $\epsilon,\delta>0$ for an alphabet of size $\omega(1)$. They also provided particular constants $\varepsilon_\sigma,\delta_\sigma$ for an alphabet of a constant size $\sigma \ge 3$ such that, under a stronger 3SUM-hardness assumption, jumbled indexing requires $\Omega(n^{2-\varepsilon_\sigma})$ preprocessing time or $\Omega(n^{1-\delta_\sigma})$ queries.
We use the techniques from both results in our algorithm and conditional lower bound for Ab-squares, respectively.
The lower bound of Amir et al.\ was later improved and extended to both existence and reporting variants and any constant $\sigma \ge 2$ by Goldstein et al.~\cite[Section 7]{DBLP:conf/esa/GoldsteinKLP16} with the aid of randomization. Moreover, recently an unconditional lower bound for the reporting variant was given in~\cite{DBLP:conf/soda/AfshaniDKN20}.

\subparagraph*{Our techniques.}
A subsequence of three distinct
positions is a 3-term double arithmetic
progression (3dap in short) if it is an
arithmetic progression and the elements on these positions also form an arithmetic progression.
The problem of finding a 3dap in a sequence is denoted by $\MidSUMthree$.
It is an odd 3dap if the first and the third
positions are odd and the middle position is even.
The corresponding problem is denoted by $\MidSUMone$.
First we reduce the convolution problem  3SUM (known to be 3SUM-hard) to the $\MidSUMthree$ problem via $\MidSUMone$ as an intermediate problem. This uses a divide-and-conquer approach and a partition of sets into sets avoiding \emph{bad} arithmetic progression of length 3.

The $\MidSUMthree$ problem reduces in a simple way
to detection of an additive square, showing that the latter problem is 3SUM hard.

Next, the $\MidSUMthree$ problem is encoded as a string. We follow the high-level idea from Amir et al. Instead of checking equality of numbers, we can check equality of their 
remainders modulo sufficiently many prime numbers. Then, each prime number corresponds to a distinct characters. If the numbers are $\mathrm{poly}\,n$ then  only $\Oh(\log n)$ prime numbers are
needed. However, there is a certain technical complication, already present in the paper of Amir et al., which needs an introduction
of additional gadgets working as \emph{equalizers}. The details, compared with construction of Amir at al., are different, mostly because in the end we want to ask about detection, not indexing. 

Then we consider the problem  of computing all centers of Ab-squares, this requires new gadgets. We show that computing all centers of Ab-squares is 3SUM-hard, as well
as detection of any Ab-square which is \emph{well centred}.

Later we extend this to detection of any odd Ab-square.  We use a construction 
of a string over the alphabet of size 4 with no Ab-square. The input string is ``shuffled'' with such a string, with some separators added. This forces odd Ab-squares to be \emph{well centered}, 
in this way we reduce the previously considered  problem of detection of any well-centred Ab-square to the detection of
any odd Ab-square. Ultimately, this shows that the latter problem is 3SUM-hard.

\section{From \texorpdfstring{$\ConvThreeSUM$}{Conv3SUM} to finding double 3-term arithmetic progressions}\label{sec:Conv3SUM->additive}
For integers $a,b$, by $[a,b]$ we denote the set $\{a,\dots,b\}$.
We use the following convolution variant of the $\ThreeSUM$ problem that is $\ThreeSUM$-hard; see \cite{DBLP:conf/soda/ChanH20,DBLP:conf/soda/KopelowitzPP16,DBLP:conf/stoc/Patrascu10} for both randomized and deterministic reductions. As already noted in \cite{DBLP:conf/icalp/AmirCLL14}, the range of elements can be made $[-N^2,N^2]$ using a randomized hashing reduction from \cite{DBLP:journals/algorithmica/BaranDP08,DBLP:conf/stoc/Patrascu10}.

\smallskip
\defproblem{$\ConvThreeSUM(\bar{x})$}{A sequence $\bar{x}=[x_1,\dots,x_N] \in [-N^2,N^2]$}
{Yes if there are $i \ne j$ such that $x_i+x_j=x_{i+j}$; no otherwise.
}

\newcommand{\I}{\mathcal{I}}
\newcommand{\MID}{\mathit{mid}}
\newcommand{\SOLVE}{\mathit{SOLVE}}

Let us denote $\MID(a,b)=(a+b)/2$ and define
 the condition: \[\MidCond_{\bar{x}}(i,j)\;=\;  (\, i \ne j\ \land\ j-i\ \text{is even}\ \land\  x_{\MID(i,j)}=\MID(x_i,x_j)\,).\]
We omit the subscript ${\bar{x}}$ if it is clear from the context. The last part of the condition is equivalent to $x_j-x_{\MID(i,j)}=x_{\MID(i,j)}-x_i$.

 Our first goal is to reduce the $\ConvThreeSUM$ problem to the following one with $K=N^{\Oh(1)}$.

\smallskip
\defproblem{Double 3-Term Arithmetic Progression, $\MidSUMthree(\bar{x}$)}{$\bar{x}=[x_1,\dots,x_n]$, each of $x_i$ is in $[0,K]$.
}{ $(\exists\, i,j)\;\MidCond(i,j)$.
}

 In \cref{sec1} we obtain a reduction of $\ConvThreeSUM$ to an intermediate version of $\MidSUMthree$ with additional constraints on $i,j$, and in \cref{sec2} we show how these constraints can be avoided.

\subsection{From \texorpdfstring{$\ConvThreeSUM$}{Conv3SUM} to \texorpdfstring{$\MidSUMone$}{MidSUM1}}\label{sec1}
Let us fix an integer sequence $x_1,\dots,x_N$. For an arithmetic progression (arithmetic sequence) $\I=i_1,\dots,i_n$, where $1 \le i_1 < \dots < i_n \le N$, i.e.\ $i_2-i_1 = \dots = i_n-i_{n-1}$, we
define the following extended functions.
\begin{align*}\ConvThreeSUM(\bar{x},\I)\,&=\,
(\,\exists\, i_a,i_b\in \I \; :\; x_{i_a}+x_{i_b}=x_{i_a+i_b},\,i_a < i_b)\\
{\OddConv}(\bar{x},\I)\,&=\,
(\,\exists\, i_a,i_b\in \I \; :\; x_{i_a}+x_{i_b}=x_{i_a+i_b},\, a+b\  \mbox{is odd}).
\end{align*}
Note that it can happen that $i_a+i_b\notin \I$. For a fixed $\bar{x}$ the  input size is $|\I|$.
\newcommand{\odd}{\mathit{odd}}
\newcommand{\even}{\mathit{even}}
 \begin{lemma}\label{lem:Conv2OddConv}
An instance of $\ConvThreeSUM(\bar{x})$ can be reduced to an alternative of $\Oh(N)$ instances of $\OddConv(\bar{x},\I)$ of total size $\Oh(N\log N)$ in $\Oh(N \log N)$ time.
\end{lemma}
\begin{proof}
If $\I=i_1,\dots,i_n$, by $\I_{\odd}$ and $\I_{\even}$ we denote the subsequences $i_1,i_3,\dots$ and $i_2,i_4,\dots$, respectively. 
We proceed recursively as shown in the following function $\ConvThreeSUM$, with the first call to $\ConvThreeSUM(\bar{x},[1,2,\ldots,N])$.

\begin{algorithm}\vskip 0.1cm
\KwSty{function} $\ConvThreeSUM(\bar{x},\I)$\vskip 0.1cm\hspace*{0.1cm}
    Comment: $\I$ is an arithmetic progression\\
    \vskip 0.2cm\hspace*{0.1cm}
    \lIf{$|\I| \le 2$}{\Return{\KwSty{false}}}\vskip 0.2cm\hspace*{0.1cm}
    \Return $\OddConv(\bar{x},\I) \lor \ConvThreeSUM(\bar{x},\I_{\odd}) \lor \ConvThreeSUM(\bar{x},\I_{\even})$\;\vskip 0.1cm
\end{algorithm}

\textbf{Correctness.}
Let $\I=i_1,\dots,i_n$ and assume there are two indices $a,b$ such that $x_{i_a}+x_{i_b}=x_{i_a+i_b}$.
If $a+b$ is odd, then $\OddConv(\bar{x},\I)$ returns true. 
Otherwise both $a,b$ are of the same parity, so
$i_a,i_b\in \I_{\odd}$ or $i_a,i_b\in \I_{\even}$. 
Consequently, the problem is split recursively
into subproblems that correspond to $\I_{\odd}$ and $\I_{\even}$.

\textbf{ Complexity.}
Let us observe that one call to $\ConvThreeSUM(\bar{x},\I)$ creates an instance of $\OddConv$ of $\Oh(|\I|)$ size in $\Oh(|\I|)$ time ($\bar{x}$ does not change). Let $\#(n)$ and $S(n)$ denote the total number and size of all instances of $\I$ generated by $\ConvThreeSUM(\bar{x},\I)$, when initially $|\I|=n$. We then have \[\#(n)=\#(\lfloor n/2\rfloor) + \#(\lceil n/2\rceil)+1\quad\mbox{and}\quad S(n)=S(\lfloor n/2\rfloor) + S(\lceil n/2\rceil) + \Theta(n),\]
which yields $\#(N)=\Oh(N)$ and $S(N)=\Oh(N \log N)$.
The reduction takes $\Oh(S(N))$ time.
\end{proof}

We say that a 3-element arithmetic progression is a \emph{good} progression if the middle element is even and two others are odd and
introduce the following problem. 

\defproblem{{ $\MidSUMone$}($\bar{x}$)}{$\bar{x}=[x_1,\dots,x_n]$, each of $x_i$ is in  $[-\Oh(N^2),\Oh(N^2)]$.
}{$(\,\exists\, i,j\,)\; [\, \MidCond(i,j)$  and $(i,\MID(i,j),j)$ is a good progression\,].
}

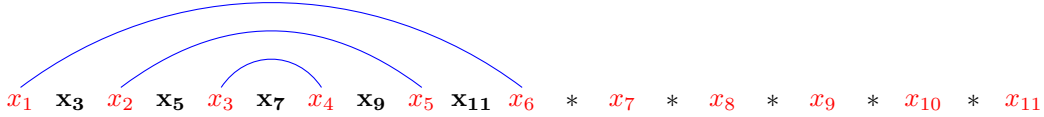
\begin{figure}[htpb]
    \centering
\begin{tikzpicture}[xscale=0.66]
    \foreach \p/\c in {
    1/\textcolor{red}{x_1}, 2/\mathbf{x_3}, 3/\textcolor{red}{x_2}, 4/\mathbf{x_5},  5/\textcolor{red}{x_3}, 6/\mathbf{x_7}, 7/\textcolor{red}{x_4}, 8/\mathbf{x_9}, 9/\textcolor{red}{x_5}, 10/\mathbf{x_{11}}, 11/\textcolor{red}{x_6}, 12/*, 13/\textcolor{red}{x_7}, 14/*, 15/\textcolor{red}{x_8}, 16/*, 17/\textcolor{red}{x_9}, 18/*, 19/\textcolor{red}{x_{10}}, 20/*, 21/\textcolor{red}{x_{11}}
 }{
    \draw (\p,0) node {\vphantom{$\mathbf{x_{11}}$}$\c$};
 }
 \begin{scope}[yshift=-0.3cm]
 \draw[blue] (5,0.5) .. controls (5.5,1) and (6.5,1) .. (7,0.5);
 \draw[blue] (3,0.5) .. controls (5,1.5) and (7,1.5) .. (9,0.5);
 \draw[blue] (1,0.5) .. controls (4,2) and (8,2) .. (11,0.5);
 \end{scope}
\end{tikzpicture}
    \caption{The sequence constructed in \cref{lem:OddConv->MidSUM1} for $\bar{x}\,=\, [x_1,x_2,x_3,\dots,x_{11}]$ and $\I=(1,2,\ldots,11)$ after the first two steps
    ($*$ denotes $\alpha_{11}$). Note that the elements connected by arcs all have their sum of indices equal to $7$; this is because $\I$ is an arithmetic progression.}
    \label{fig:birds}
\end{figure}

\begin{lemma}\label{lem:OddConv->MidSUM1}
$\OddConv(\bar{x},\I)$ is reducible in $\Oh(|\I|)$ time and space to $\MidSUMone(\bar{y})$, where $|\bar{y}|=\Oh(|\I|)$.
\end{lemma}
\begin{proof} Let $\I=i_1,\dots,i_n$. Define $\alpha_N=2N^2+1$ and let
$\bar{y}$ be a sequence of length $2n-1$ that is created as follows:
\begin{enumerate}
    \item put $x_{i_1},x_{i_2},\dots,x_{i_n}$ at subsequent odd positions in $\bar{y}$;
    \smallskip
    \item at each even position $2j$, put $x_{i_j+i_{j+1}}$ or, if $i_j+i_{j+1}>N$, put $\alpha_N$.\smallskip
    \item multiply elements on even positions by 2.
\end{enumerate}

After the first two steps $\OddConv(\bar{x},\I)$ is equivalent to $(\exists i,j)\;y_{\MID(i,j)}=y_i+y_j$ for odd $i,j$ and even $\MID(i,j)$; see~\cref{fig:birds}.
Then, after the third step, $\OddConv(\bar{x},\I)$ is equivalent to $\MidSUMone(\bar{y})$.
\end{proof}
\vspace*{-0.8cm}

\subsection{From \texorpdfstring{$\MidSUMone$}{MidSUM1} to \texorpdfstring{$\MidSUMthree$}{MidSUM}}\label{sec2}

Our main tool in this subsection is partitioning a set of integers into
progression-free sets.
A set of integers $A$ is called \emph{progression-free} if it does not contain a non-constant three-element arithmetic progression. We use the following recent result that extends a classical paper of Behrend~\cite{Behrend331}.
\begin{theorem}[\cite{DBLP:conf/stoc/0001GS20}]\label{thm:Behrend}
Any set $A \subseteq [1,n]$ can be partitioned into $n^{o(1)}$
progression-free sets in $n^{1+o(1)}$ time.
\end{theorem}
\newcommand{\F}{\mathcal{F}}
\newcommand{\RED}{\mathsf{RED}}
\newcommand{\BLUE}{\mathsf{BLUE}}
\newcommand{\GRAY}{\mathsf{GREEN}}
\begin{lemma} \label{lem:family} We can construct in $n^{1+o(1)}$   time a  family
$\F$ of  $n^{o(1)}$ subsets of $[1,n]$ satisfying:
\begin{enumerate}[(a)]
\item\label{ita} Each good 3-element progression is contained in some
$S\in \F$.
\item\label{itb} If $ S\in \F$, then all 3-element arithmetic progressions in $S$ are
good.
\end{enumerate}
\end{lemma}
\begin{proof}
Let us divide the elements from $[1,n]$ into three classes:
\begin{align*}
    \BLUE&=\{i\le n\,:\, i \text{ is even}\,\},\\
    \RED&=\{i\le n\,:\, i\bmod 4=1\}, \quad
    \GRAY=\{i\le n\,:\, i\bmod 4=3\}.
\end{align*}
Each element $i\in[1,n]$ has the colour blue, red or green of its corresponding class.
Each class forms an arithmetic progression.

\smallskip \noindent  A progression is called \emph{multi-chromatic}
if its elements are of three distinct colours.
Let us observe that 
 a 3-element progression is good if and only if it is multi-chromatic.
 Indeed, this is because if $i,j \in \RED$ (or $\GRAY$), then $\MID(i,j)$ is odd.
 
Now instead of good progressions we will deal with multi-chromatic progressions.
We treat sets of integers as increasing sequences and for a set $C=\{c_1,\dots,c_m\}$ we denote by $C_{\odd}$ and $C_{\even}$ the subsets $\{c_1,c_3,\dots\}$ and $\{c_2,c_4,\dots\}$.

For example 
$\BLUE_{odd}=\{i\le n\,:\, i\bmod 4=2\}, \ \RED_{even}=\{i\le n\,:\, i\bmod 8=5\}$.

Our construction works as follows:

\begin{enumerate}
    \item Partition the set $[1,n]$ into classes $\BLUE,\RED,\GRAY$.
    \item For each class $C\in \{\BLUE,\RED,\GRAY\}$  partition it in $n^{1+o(1)}$ time into a family $\F_{C}$ of $n^{o(1)}$ progression-free sets with the use of Theorem \ref{thm:Behrend}.
    \item Refine each partition $\F_C$, splitting each set $X\in \F_C$ into two sets $X\cap C_{\odd}$,
    $X\cap C_{\even}$, so that for each set $X$ in the new refined partition $\F_C$ we have $X\subseteq C_{\odd}$ or $X\subseteq C_{\even}$. Each family is still of size $n^{o(1)}$.
    \item Return $\F\,=\, \{\,X\cup Y\cup Z\;:\;X\in \F_{\BLUE}, Y\in \F_{\RED}, Z\in \F_{\GRAY}\,\}$.
 \end{enumerate}   
    % ---------------------------------------------------
% ---------------------------------------------------

\textbf{ Proof of point (\ref{ita})}.
Each multi-chromatic progression is contained in some $S\in \F$ since each element of $C$ is contained in a set from $\F_C$.

\smallskip
\textbf{ Proof of point (\ref{itb})}.
The proof is by contradiction. Assume that $S \in \F$ contains a progression which is not multi-chromatic. There are two cases.
\begin{description}
\item \textbf{ Case 1:} the progression is monochromatic, hence it appears in a single set $X\in \F_C$.
However every $X$ is progression-free (step 2), hence such a progression cannot appear in any $S\in \F$; a contradiction.
\item \textbf{ Case 2:} the progression contains exactly two different colors.
Observe that 
if $i  \bmod p= \MID(i,j) \bmod p=r$, then $j \bmod p = r$ (if the middle element of progression belongs to the same class as one of the other elements, then the triple is monochromatic),
hence the two-coloured arithmetic progression has to consist of $i,j\in C$ and $\MID(i,j)\notin C$.

Since $i,j$ both belong to $C_{\odd}$ or $C_{\even}$ (step 3), $\MID(i,j)$ must belong to $C$ (if $i\bmod 2p=j\bmod2p$, then $i\bmod p=\MID(i,j)\bmod p$). Consequently, the progression cannot contain exactly two colours; a contradiction.\qedhere
\end{description} 
\end{proof}

\newcommand{\restr}{\mathit{restr}}
Our next tool is a \emph{deactivation} of a set of elements which indexes are
not in a given set $E$, that is, omitting them in the computation of a solution.
For  $E\subseteq [1,n]$ the operation $\restr(\bar{x},E)$
replaces each element $x_i$ on position $i \notin E$ by
$5\max\{\mathit{MAX},n^2\}+i^2$, where $\mathit{MAX}=\max_{k\ge 1}\, |x_k|$.
\begin{lemma}\label{April11} $
\MidSUMthree(\restr(\bar{x},E))
\iff (\exists\, i,j)\;\MidCond_{\bar{x}}(i,j)\; \land\ i,j,\MID(i,j)\in E$.
\end{lemma}
\begin{proof}
The $(\Leftarrow)$ part if obvious, so it suffices to show $(\Rightarrow)$.
If at least one, but not all, of $i,j,\MID(i,j)$ is not in $E$, then it can be checked that $\MidCond_{\bar{y}}(i,j)$ cannot hold for $\bar{y}=\restr(\bar{x},E)$ because $y_{\MID(i,j)}$ and $\MID(y_i,y_j)$ differ by at least $M:=\max\{\max_k\{|x_k|\},n^2\}$. Indeed, there are seven possible cases:
\begin{enumerate}
    \item $i,\MID(i,j),j\notin E$, then $\MID(y_i,y_j)-y_{\MID(i,j)}=\frac{i^2+j^2}{2}-(\frac{i+j}{2})^2=(\frac{i-j}{2})^2>0\ \mbox{since}\ i\neq j$
    \item $i\in E$, $\MID(i,j),j\notin E$, then
    $\MID(y_i,y_j)-y_{\MID(i,j)}= -\frac{5}{2}M + \frac{j^2}{2} -(\frac{i+j}{2})^2 +\frac{x_i}{2}\le -2M+\frac{x_i}{2}\le -M$
    \item $j\in E$, $i,\MID(i,j)\notin E$ works as the previous case
    \item $\MID(i,j)\in E$, $i,j\notin E$, then 
    $\MID(y_i,y_j)-y_{\MID(i,j)}=5M+\frac{i^2+j^2}{2}-x_{\MID(i,j)}\ge 4M$
    \item $i,\MID(i,j)\in E$, $j\notin E$, then
    $\MID(y_i,y_j)-y_{\MID(i,j)}=\frac{5}{2}M+\frac{j^2}{2}+\frac{x_i}{2}-x_{\MID(i,j)}\ge M$
    \item $\MID(i,j),j\in E$, $i\notin E$ works as the previous case
    \item $i,j\in E$, $\MID(i,j)\notin E$, then
    $\MID(y_i,y_j)-y_{\MID(i,j)}=\frac{x_i+x_j}{2}-5M-(\frac{i+j}{2})^2\le -3M$.
\end{enumerate}
Hence, apart from the first case, where none of the indices belongs to $E$, the absolute value of difference is at least $M$.

Otherwise, if all the positions $i,j,\MID(i,j)$ are not in $E$, then $\MidCond_{\bar{y}}(i,j)$ does not hold because $\MID(i^2,j^2)-(\frac{i+j}{2})^2=(\frac{i-j}{2})^2>0\ \mbox{since}\ i\neq j.$
\end{proof}
An instance $\bar{x}$ is called an \textbf{odd-half} instance if $\MidCond(i,j)$ is false for $i,j$ such that
$(j-i)/2$ is even (equivalently, for $i,j$ such that $i$ and $\MID(i,j)$ have the same parity).
Efficient equivalence 
\[\MidSUMone(\bar{x})\ \Leftrightarrow\ (\exists\,S\in \F\,) \; \MidSUMthree(\restr(\bar{x},S))\] follows now from \cref{lem:family,April11}. 

This  produces only odd-half instances because only good progressions are left in the construction of \cref{lem:family}.
The instances have elements in $[-\Oh(N^2),\Oh(N^2)]$. We can increase all the elements by $\Oh(N^2)$ so that they become non-negative.
This implies:

\begin{lemma}\label{lem:MidSUM1->3}
An instance of $\MidSUMone$ can be reduced in $n^{1+o(1)}$ time to $n^{o(1)}$ odd-half instances of $\MidSUMthree$ of total size $n^{1+o(1)}$ and with elements up to $K=\Oh(N^2)$.
\end{lemma}

\noindent
Finally, we show that the resulting instances can be glued together to a single equivalent one.

\begin{theorem}\label{Conv->Mid}
An instance of $\ConvThreeSUM$ can be reduced in $N^{1+o(1)}$ time to an odd-half instance of $\MidSUMthree$ of size $n=N^{1+o(1)}$ with elements up to $K=N^{3+o(1)}$.
\end{theorem}
\begin{proof}
With \cref{lem:Conv2OddConv,lem:OddConv->MidSUM1,lem:MidSUM1->3} we obtain a reduction from $\ConvThreeSUM$ to $N^{1+o(1)}$ odd-half instances of $\MidSUMthree$ of total size $N^{1+o(1)}$. The instances have elements in $[0,\Oh(N^2)]$. We will show that these instances can be reduced to a single odd-half instance of $\MidSUMthree$ of size $N^{1+o(1)}$ with elements in the range $[0,N^{3+o(1)}]$ in time $N^{1+o(1)}$. The resulting instance will return true if and only if at least one of the input instances does.

Let $t=N^{1+o(1)}$ be the number of the instances of $\MidSUMthree$, numbered $1$ through $t$. We use \cref{thm:Behrend} \footnote{Actually, a deterministic version of Behrend's construction from \cite{DBLP:conf/stoc/0001GS20} or an earlier construction of Salem and Spencer~\cite{Salem561} would suffice here.} and pick the largest constructed progression-free set $A\subseteq [1,m]$, for some $m$. By the pigeonhole principle, $|A| \ge m^{1-o(1)}$. We select $m$ that is large enough so that $m^{1-o(1)} \ge t$, so $m=t^{1+o(1)}=N^{1+o(1)}$, and trim the set $A$ to the size $t$.
Let $A=\{a_1,\dots,a_t\}$.
For instance $i$ we multiply all its elements by $2m$ and add to each element the value $a_i$.
Finally we concatenate all the instances.

If any of the input instances returns true, then so does the output instance, since multiplication by and addition of the same number to all elements cannot affect the outcome of a single instance.
If none of the input instances returns true, then the only possibility for the output instance to return true is to contain a 3-element arithmetic progression with elements from multiple parts corresponding to the input instances. However, this is impossible since, taken modulo $2m$, the progression would form an arithmetic progression in the set $A$.
\end{proof}
\begin{corollary}
The general $\MidSUMthree$ problem is also $\ThreeSUM$-hard.
\end{corollary}
\begin{remark}
Similarly as in $\ConvThreeSUM$, techniques from \cite{DBLP:journals/algorithmica/BaranDP08} can be used to hash down the range in $\MidSUMthree$ to integers of magnitude $\Oh(N^2)$ (cf.\ \cite{DBLP:conf/icalp/AmirCLL14}), using randomization.
\end{remark}
\begin{remark}
The AVERAGE problem (introduced by J.\, Erickson~\cite{Erickson}) asks if there are distinct elements $a,b,c \in S$ such that $a+b=2c$ for a given set $S$ of $n$ integers. It was recently shown to be $\ThreeSUM$-hard~\cite{DBLP:conf/stoc/0001GS20}. The $\MidSUMthree$ problem can be viewed as a convolution version of the AVERAGE problem\footnote{\url{https://cs.stackexchange.com/questions/10681/is-detecting-doubly-arithmetic-progressions-3sum-hard/10725\#10725}}. The ideas based on almost linear hashing used in the reductions from $\ThreeSUM$ to $\ConvThreeSUM$ \cite{DBLP:conf/stoc/Patrascu10,DBLP:conf/soda/ChanH20} can be extended with some effort to reduce AVERAGE to $\MidSUMthree$.
We presented a different reduction that additionally directly leads to an instance of $\MidSUMthree$ with an odd-half property,  
which is essential in our proof of $\ThreeSUM$-hardness of computing Ab-squares (see the proof of \cref{lem:well-placed}).
\end{remark}

\subsection{Hardness of detecting additive squares}
If the alphabet is a set of integers, then a string $W$ is called an \emph{additive square} if
$W=UV$, where $|U|=|V|$ and $\sum_{i=1}^{|U|}\, U[i] = \sum_{i=1}^{|V|}\, V[i]$.

\begin{theorem}\label{thm:additive_sq}
Finding an additive square in a length-$N$ sequence composed of integers of magnitude $N^{\Oh(1)}$ is $\ThreeSUM$-hard.
\end{theorem}
\begin{proof}
We use \cref{Conv->Mid} to reduce $\ConvThreeSUM$ to an instance of $\MidSUMthree$ of size $n=N^{1+o(1)}$ with elements in the requested range. $\MidSUMthree$ returns true on an instance $x_1,\dots,x_n$ if and only if the sequence $x_2-x_1,x_3-x_2,\dots,x_n-x_{n-1}$ contains an additive square. As the reduction works in $N^{1+o(1)}$ total time, the conclusion follows.
\end{proof}

\section{From arithmetics to Abelian stringology}
We use capital letters to denote strings and lower case Greek letters to denote sets of integers. We assume that the positions in a string $S$ are numbered 1 through $|S|$, where $|S|$ denotes the length of $S$. By $S[i]$ and $S[i..j]$ we denote the $i$th letter of $S$ and the string $S[i] \cdots S[j]$ called a factor of $S$. The reverse of string $S$, i.e.\ the string $S[|S|] \cdots S[1]$, is denoted as $S^R$. By $\varepsilon$ we denote the empty string. By $\Alphabet(S)$ we denote the set of distinct letters in $S$.

We denote Ab-equivalence of $U$ and $V$ by $U\cong V$. For a string $U$, by $\Parikh(U)$ we denote the Parikh vector of $U$. Then $U \cong V$ if and only if $\Alphabet(U)=\Alphabet(V)$ and $\Parikh(U)=\Parikh(V)$.

We use an encoding of Amir et al.~\cite{DBLP:conf/icalp/AmirCLL14} based on the Chinese remainder theorem to connect $\ConvThreeSUM$-type problems with Abelian stringology.

Let $p_1<p_2<\dots<p_k$ be prime numbers.
The Chinese remainder theorem states that if one knows the remainders $r_1,r_2,\dots,r_k$ of an integer $x$, such that $0\le x<\prod\, p_i$, when dividing by $p_i$'s, then one can uniquely determine $x$. 
Assuming that the remainders of an integer  $x$ are $r_1,r_2,\dots,r_k$, we could encode $x$ as a possibly short
string $\aa_1^{r_1}\aa_2^{r_2}\cdots \aa_k^{r_k}$ over an alphabet $\{\aa_1,\aa_2,\ldots, \aa_k\}$
(the symbols correspond to consecutive prime numbers). 

For example for primes 2,3,5 the encoding of 11 would be $\aa_1^1\aa_2^2\aa_3^1$
since its remainders modulo 2,3,5 are 1,2,1, respectively. However, we are interested in encodings of 
subtractions of one number from another one, and it is more complicated.

\newcommand{\EXP}{\mathit{EXP}}

Let $\bar{x}=[x_1,\dots,x_n]$ be an instance of $\MidSUMthree$ and $r_1^{(i)},r_2^{(i)},\dots,r_k^{(i)}$ be remainders of $x_i$ modulo $p_1,p_2,\ldots, p_k$.
Like Amir et al.~\cite{DBLP:conf/icalp/AmirCLL14}, we define for $1 \le i<n$ and $1 \le j\le k$, \[ \EXP_i(j)=r_j^{(i+1)} - r_j^{(i)} + d\text{ where }d=\max_{j=1}^k p_j,\ \ 
\SS_i\;=\;\aa_1^{\EXP_i(1)}\, \aa_2^{\EXP_i(2)} \cdots \aa_k^{\EXP_i(k)}.\]
We choose a sequence $p_1,\dots,p_k$ of $k$ distinct primes such that $p_1 \cdots p_k > \max\{x_i\}$.
In this way we encode the difference $x_j-x_i$, for $j>i$, by a string $\SS_i\,\SS_{i+1}\cdots \SS_{j-1}$.
An obstacle is the potentially possible inequality
$(a \bmod p)-(b \bmod p)\ne (a-b) \bmod p$.
For example $$(4\ \text{mod}\ 3)-(2\ \text{mod}\ 3)\,=\,  ((4-2)\ \text{mod}\ 3)\, - \,3.$$

However a small correction is sufficient, due to the following observation.
\begin{observation} $(a \bmod p)-(b \bmod p)+q\,=\, (a-b) \bmod p$, where
$q\in  \{0,p\}$.
\end{observation}

\noindent
If we apply the encoding to an instance $\bar{x}=x_1,\dots,x_n$ of $\MidSUMthree$, we obtain a lemma that is analogous to \cite[Lemma 1]{DBLP:conf/icalp/AmirCLL14}.

\begin{lemma}\label{lem1} $\MidCond(i,j)$ holds for $i<j$, $j-i$ even, iff for each $t\in [1,k] $, 
there are 
$e_t,f_t\in\{0,p_t\}$, such that 
\[ e_t+\EXP_i(t)+\EXP_{i+1}(t)+\dots+ \EXP_{\MID(i,j)-1}(t)
\,=\,\]
\[\hspace*{1cm} \EXP_{\MID(i,j)}(t)+\EXP_{\MID(i,j)+1}(t)+\cdots+\EXP_{j-1}(t)+f_t.\]
\end{lemma}

\medskip
Let $\Psi$ be a morphism such that $\Psi(i)=\aa_i^{p_i}$ for each $i=1,\dots,k$.
We  treat a set $U=\{u_1,\dots,u_w\}$ as a string $u_1\cdots u_w$, where $u_1<u_2<\ldots <u_w$. If we interpret the vector 
$(\EXP_i(1),\EXP_i(2),\dots,\EXP_i(k))$ as $\SS_i$, then Lemma~\ref{lem1}
directly implies the following fact.

\begin{restatable}{lem}{slem}\label{March26} Assume $i<j$ and $j-i$ is even. 
Then 
 \[\MidCond(i,j)\ \iff\ (\, \Psi(\alpha) \,\SS_i\SS_{i+1}\cdots \SS_{\MID(i,j)-1}\cong \SS_{\MID(i,j)}\cdots \SS_{j-1}\; \Psi(\beta)\,)\]
for some disjoint subsets $\alpha, \beta$ of $[1,k]$.
\end{restatable}

\section{Hardness of computing all centers of Ab-squares}\label{subsec:centers}
We construct a text $T$ over the alphabet $\{\aa_1,\dots,\aa_k,\bb,\xx,\yy,\#,\$\}$ such that $\MidSUMthree$ has a solution if and only if $T$ contains an Ab-square with one of specified centers, so-called \emph{well-placed} Ab-square.

First we extend each $\SS_i$ to have the same length $M \ge \max_{i=1}^{n-1} |\SS_i|$, to be defined later.
Intuitively, it is needed to control the number of $\SS_i$'s in the strings from \cref{March26}.
We append $M-|\SS_i|$ occurrences of a letter $\bb$ to each $\SS_i$. Let $\SS^I_i$ denote this modified string. 

Lemma~\ref{March26} immediately implies the following fact.

\begin{lemma}\label{March24_2}
Assume $i<j$ and $j-i$ is even. 
Then   
\[\MidCond(i,j)\ \iff\ (\, \bb^e\Psi(\alpha) \,\SS_i^I\SS_{i+1}^I\cdots\SS_{\MID(i,j)-1}^I\,\cong  \SS_{\MID(i,j)}^I\cdots \SS_{j-1}^I \Psi(\beta)\bb^f\, ),\]
for some disjoint subsets $\alpha, \beta$ of $[1,k]$,
where $e+|\Psi(\alpha)|=f+|\Psi(\beta)|$ with $\min(e,f)=0$.
\end{lemma}
The parts $\bb^e\Psi(\alpha)$, $\Psi(\beta)\bb^f$ in the above lemma can be treated as \emph{equalizers}.
Let us note that in the above lemma we can assume that $\max(e,f) \le \max(|\Psi(\alpha)|,|\Psi(\beta)|) \le kd$.

A pair of disjoint sets $\alpha,\beta$ that satisfies $\alpha \cup \beta=[1,k]$ will be called a \emph{2-partition} of $[1,k]$.
For a 2-partition $(\alpha,\beta)$ of $[1,k]$, we use the string \[\Gamma(\alpha,\beta)=\#\,\alpha\,\$\,\bb^{kd}\,\#\,\beta\,\$,\] called a \emph{$\Gamma$-string}. 
If $k=4$, $d=7$, an example of a $\Gamma$-string is
$\Gamma(2,\, 1\,3\,4)\,=\,\#\,2\;\$\;\bb^{28}\;\#\;1\,3\,4\;\$$.

Let $(\pi_1,\pi'_1),(\pi_2,\pi'_2),\dots,(\pi_m,\pi'_m)$
be the sequence of all $m=4^k$ pairs of $\Gamma$-strings. Define
\[U=\pi_m\pi_{m-1}\ldots \pi_1,\ \ V=\pi'_1\,\pi'_2\,\ldots \pi'_m.\]
We have $\{\pi_1,\dots,\pi_m\}=\{\pi'_1,\dots,\pi'_m\}$, so $U \cong V$.

\begin{observation}\label{March21}
For disjoint subsets $\alpha,\beta \subseteq [1,k]$ and integers $0 \le e,f\le kd$, there are decompositions
$U\;=\; U_1\,\bb^e\,\#\,\beta\,\$\,U_2$ and 
$V\;=\; V_1\,\#\,\alpha\,\$\,\bb^f\,V_2$, 
where $U_2\cong V_1$.
\end{observation}
Let us recall the morphism $\Psi$ such that $\Psi(i)=\aa_i^{p_i}$ for each $i\in[1,k]$. We define additionally  $\Psi(c)=c$ for $c \in \{\bb,\#,\$\}$ and set
    \[\BB\,=\,\Psi(U),\ \AA\,=\,\Psi(V),\ \ M=|\AA|=|\BB|.\]
Let us observe that indeed $\max_{i=1}^{n-1} |\SS_i| \le M$ holds since $|\SS_i| \le kd+\sum_{j=1}^k p_j$ and the length of $\Psi(W)$ for any $\Gamma$-string $W$ is $kd+\sum_{j=1}^k p_j+4$. 

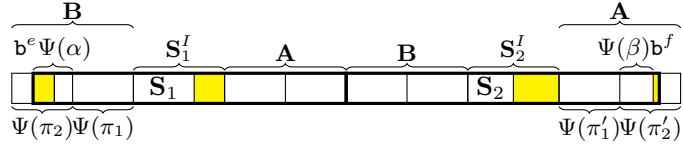
\begin{figure}[th]
\centering
\begin{tikzpicture}[scale=0.4]
    \begin{scope}
      \draw[] (0,0) rectangle (4,1);
      \draw (2,0) -- (2,1);
      \filldraw[yellow] (0.7,0) rectangle (1.4,1);
      \draw (1.4,0) -- (1.4,1);
      \draw[snake=brace] (0.7,1.1) -- node[above] {\small $\bb^e \Psi(\alpha)$} (2,1.1);
      \draw (1.4,0) -- (1.4,1);
      \draw[snake=brace] (2,-0.1) -- node[below] {\small $\Psi(\pi_2)$} (0,-0.1);
      \draw[snake=brace] (4,-0.1) -- node[below] {\small $\Psi(\pi_1)$} (2,-0.1);
      \draw[snake=brace] (0,2.5) -- node[above] {\small $\BB$} (4,2.5);
    \end{scope}
    \begin{scope}[xshift=4cm]
      \filldraw[yellow] (2,0) rectangle (3,1);
      \draw[] (0,0) rectangle (3,1);
      \draw (1,0.5) node {$\SS_1$};
      \draw (2,0) -- (2,1);
      \draw[snake=brace] (0,1.1) -- node[above] {\small $\SS_1^I$} (3,1.1);
    \end{scope}
    \begin{scope}[xshift=7cm]
      \draw[] (0,0) rectangle (4,1);
      \draw (2,0) -- (2,1);
      \draw[snake=brace] (0,1.1) -- node[above] {\small  $\AA$} (4,1.1);
    \end{scope}
    \begin{scope}[xshift=11cm]
      \draw[] (0,0) rectangle (4,1);
      \draw (2,0) -- (2,1);
      \draw[snake=brace] (0,1.1) -- node[above] {\small $\BB$} (4,1.1);
    \end{scope}
    \begin{scope}[xshift=15cm]
      \filldraw[yellow] (1.5,0) rectangle (3,1);
      \draw[] (0,0) rectangle (3,1);
      \draw (0.75,0.5) node {$\SS_2$};
      \draw (1.5,0) -- (1.5,1);
      \draw[snake=brace] (0,1.1) -- node[above] {\small $\SS_2^I$} (3,1.1);
    \end{scope}
    \begin{scope}[xshift=18cm]
      \draw[] (0,0) rectangle (4,1);
      \draw (2,0) -- (2,1);
      \filldraw[yellow] (3.1,0) rectangle (3.3,1);
      \draw (3.3,0) -- (3.3,1);
      \draw[snake=brace] (2,1.1) -- node[above] {\small $\Psi(\beta)\bb^f$} (3.1,1.1);
      \draw (3.1,0) -- (3.1,1);
      \draw[snake=brace] (2,-0.1) -- node[below] {\small $\Psi(\pi'_1)$} (0,-0.1);
      \draw[snake=brace] (4,-0.1) -- node[below] {\small $\Psi(\pi'_2)$} (2,-0.1);
      \draw[snake=brace] (0,2.5) -- node[above] {\small $\AA$} (4,2.5);
    \end{scope}
    \draw[very thick] (0.7,0) rectangle (11,1);
    \draw[very thick] (11,0) rectangle (21.3,1);
\end{tikzpicture}
\caption{Internal structure of an Ab-square, shown in a thick box (proportions are symbolic), in $\BB\SS_1^I\AA\,\BB\SS_2^I\AA$.
Here $x_{\MID(1,3)}=\MID(x_1,x_3)$ and $\bb^e\Psi(\alpha)$, $\Psi(\beta)\bb^f$ are \emph{equalizers}.}
\end{figure}

We add two new letters $\xx,\yy$ and define the following string
(the symbols ``$\textcolor{violet}{\stackrel{center}{\downarrow}}$'' are not parts of the string, but only show supposed centers of Ab-squares). 
\newcommand{\NIC}{}
\begin{equation}\label{eqT}
    T\;=\;\xx\,\NIC\;\BB\; \yy\,\NIC\;\SS^I_1\; \AA\;\xx\,\NIC\textcolor{violet}{\stackrel{center}{\downarrow}}%
    \yy\,\NIC\;\BB\;  \xx\,\NIC\;\SS^I_2\; \;\AA\;\yy\,\NIC\textcolor{violet}{\stackrel{center}{\downarrow}}%
    \xx\,\NIC\;\BB\; \yy\,\NIC\;\SS^I_3\; \AA\;\xx\,\NIC\textcolor{violet}{\stackrel{center}{\downarrow}}%
    \yy\,\NIC\;\BB\; \xx\,\NIC\;\SS^I_4\; \AA\,\yy\,\NIC \cdots.
\end{equation}

An Ab-square is called \emph{well-placed} if its center is between the letters $\xx,\yy $
in any order.
Recall that, due to Theorem~\ref{Conv->Mid}, we can assume that 
 the input to $\MidSUMthree$ guarantees that only odd-half instances could have solutions.

\begin{lemma}\label{lem:well-placed}
Assume $\bar{x}$ is an odd-half instance. Then $\MidSUMthree(\bar{x})$ has  a solution if and only if $T$ contains a well-placed Ab-square.
\end{lemma}
\begin{proof}
Let $\bar{x}=[x_1,\dots,x_n]$ be an odd-half instance of $\MidSUMthree$.
We show two implications.

\smallskip
$(\mathbf{\Rightarrow})$ 

Assume that $\MidCond(i,j)$ holds for $\bar{x}$. \cref{March24_2} implies that for strings $W,Z$ such that $W \cong Z$ we have
\begin{multline}\label{eqsq}
\bb^e\#\Psi(\alpha)\$ W\,
\yy\SS_i^I \AA \xx\,\,\,
\yy\BB \xx\SS_{i+1}^I \AA \yy\,\,\,
\cdots\,\,\,
\xx\BB \yy\SS_{\MID(i,j)-1}^I \AA \xx\,\,\,
\cong\\
\quad\quad \yy\BB \xx\SS_{\MID(i,j)}^I\AA \yy\,\,\,
\cdots\,\,\,
\xx\BB \yy\SS_{j-2}^I \AA \xx\,\,\,
\yy\BB \xx\SS_{j-1}^I\,
Z \#\Psi(\beta)\$\bb^f
\end{multline}
for some disjoint subsets $\alpha, \beta$ of $[1,k]$,
where $e+|\Psi(\alpha)|=f+|\Psi(\beta)|$ with $\min(e,f)=0$.
Indeed, we use the fact that $\AA \cong \BB$ and the counts of letters $\xx$ and $\yy$ on both hand sides are equal (because $(j-i)/2$ is odd).
By \cref{March21}, we obtain a well-placed Ab-square in $T$ (or we obtain it after exchanging all letters $\xx$ with $\yy$).

\smallskip
$(\mathbf{\Leftarrow})$ Assume that $T$ has a well-placed Ab-square factor with center immediately after $\xx\BB \yy\SS_{t}^I \AA \xx$ (the case that it is immediately after $\yy\BB \xx\SS_{t}^I \AA \yy$ is symmetric). Let us investigate what can be the position $s$ of the first letter of this Ab-square.

Recall that $|\SS_i^I|=|\AA|=|\BB|=M$ for each $i \in [1,n-1]$, so $T$ can be seen as composed of blocks of length $M'=M+1$.
We will check which of these blocks can contain $s$, by checking the counts of each of the letters $\xx,\yy$ in both halves of the Ab-square. The positions of letters $\xx,\yy$ in $T$ repeat with period $6(M+1)$, so it is sufficient to inspect the first 6 blocks on each side, as the remaining ones will behave periodically; see \cref{fig:xy,fig:prow}.

\begin{figure}[htpb]
    \centering
\begin{tikzpicture}[xscale=1.19]
    \draw[thick] (5.5,-0.4) -- (5.5,0.4);
    \foreach \x/\s in {
        0/\yy\,\BB,1/\xx\,\SS_{t-1}^I,2/{\AA}\,\yy,
        3/\xx\,\BB,4/\yy\,\SS_{t}^I,5/{\AA}\,\xx,
        6/\yy\,\BB,7/\xx\,\SS_{t+1}^I,8/{\AA}\,\yy,
        9/\xx\,\BB,10/\yy\,\SS_{t+2}^I,11/{\AA}\,\xx
    }{
        \draw (\x,0) node {\vphantom{$\xx\zz\SS_{t-1}^I$} $\s$};
    }
    \foreach \x/\s in {
        0/first,1/none,2/any,3/not first,4/first,5/none
    }{
        \draw (\x,0.4) node[above] {\vphantom{ty}\textcolor{violet}{\s}};
    }
\end{tikzpicture}
    \caption{Which position in a block of $T$ can be the starting position of a well-placed Ab-square with the designated center, just counting letters $\xx,\yy$. 
    }
    \label{fig:xy}
\end{figure}
By counting letters $\xx,\yy$ in both halves of the Ab-square, it can be readily verified that $s$ cannot be in any block $\AA \xx$ or $\xx\SS_i^I$;
if in any block $\yy\BB$ or $\yy\SS_i^I$, it can only be the first position of the block; it cannot be the first position in a block $\xx\BB$; and 
it can be in any position in a block $\AA \yy$.

Moreover, $s$ cannot be the first position in a block $\yy\BB$, since this would imply, by \cref{March24_2}, that $\MidCond(i,j)$ holds for $i$ such that the block $\xx\SS_i^I$ immediately follows the $\yy\BB$ block and $j=2t-i$. However, in this case $(j-i)/2$ is even, which is impossible.

If $s$ is the first position of a block $\yy\SS_i^I$, then this implies, again by \cref{March24_2}, that $\MidCond(i,j)$ holds for $j=2t-i$. In this case $(j-i)/2$ is odd, so this is a valid solution to the corresponding $\MidSUMthree$ instance.

We are left with the case that $s$ belongs to a block $\AA \yy$ or $\xx\BB$ (and in case of $\xx\BB$ does not coincide with the position of the letter $\xx$). Henceforth it suffices to count letters different from $\xx,\yy$ in the halves.
Each of the gadgets $\AA,\BB$ is a concatenation of $m$ Ab-equivalent strings of the form $\#\,\Psi(\alpha)\,\$\,\bb^{kd}\,\#\,\Psi(\beta)\,\$$, where $\Psi(\alpha),\Psi(\beta)$ are composed of letters $\aa_i$ only. By counting the letters \# and \$ in both halves of the Ab-square, we see that $s$ can only be a position which holds the letter $\bb$ or \#. 

Hence, the Ab-square is necessarily of the form \eqref{eqsq}, which, by \cref{March24_2}, implies that $\MidCond(i,j)$ holds, where $\yy\SS_i^I$ is the first such block after the position $s$ and $j=2t-i$.
\end{proof}

\begin{figure}[h!]
    \centering
\begin{scaletikzpicturetowidth}{\textwidth}
\begin{tikzpicture}[scale=\tikzscale]
	\path[draw]	(0.0, 0.0)	grid (20.0, 1.0);
	\path[draw, fill=red!50!white]	(2.0, 0.0)	rectangle (3.0, 1.0);
	\path[draw, fill=green!50!white]	(5.0, 0.0)	rectangle (6.0, 1.0);
	\path[draw, fill=blue!50!white]	(8.0, 0.0)	rectangle (9.0, 1.0);
	\path[draw, fill=cyan!50!white]	(11.0, 0.0)	rectangle (12.0, 1.0);
	\path[draw, fill=magenta!50!white]	(14.0, 0.0)	rectangle (15.0, 1.0);
	\path[draw, fill=yellow!50!white]	(17.0, 0.0)	rectangle (18.0, 1.0);
	\path[]	(1.0, 1.0)	-- (1.0, 1.3)	node[right] () {$\bullet$};
	\path[]	(19.0, 1.0)	-- (19.0, 1.3)	node[left] () {$\star$};
	\path[]	(4.0, 1.0)	-- (4.0, 1.3)	node[left] () {$\bullet$};
	\path[]	(4.0, 1.0)	-- (4.0, 1.3)	node[right] () {$\star$};
	\path[]	(7.0, 1.0)	-- (7.0, 1.3)	node[left] () {$\star$};
	\path[]	(7.0, 1.0)	-- (7.0, 1.3)	node[right] () {$\bullet$};
	\path[]	(10.0, 1.0)	-- (10.0, 1.3)	node[left] () {$\bullet$};
	\path[]	(10.0, 1.0)	-- (10.0, 1.3)	node[right] () {$\star$};
	\path[]	(13.0, 1.0)	-- (13.0, 1.3)	node[left] () {$\star$};
	\path[]	(13.0, 1.0)	-- (13.0, 1.3)	node[right] () {$\bullet$};
	\path[]	(16.0, 1.0)	-- (16.0, 1.3)	node[left] () {$\bullet$};
	\path[]	(16.0, 1.0)	-- (16.0, 1.3)	node[right] () {$\star$};
	\path[]	(2.0, 0.0)	-- (2.0, -0.3)	node[right] () {$\star$};
	\path[]	(5.0, 0.0)	-- (5.0, -0.3)	node[right] () {$\bullet$};
	\path[]	(8.0, 0.0)	-- (8.0, -0.3)	node[right] () {$\star$};
	\path[]	(11.0, 0.0)	-- (11.0, -0.3)	node[right] () {$\bullet$};
	\path[]	(14.0, 0.0)	-- (14.0, -0.3)	node[right] () {$\star$};
	\path[]	(17.0, 0.0)	-- (17.0, -0.3)	node[right] () {$\bullet$};
	\path[draw]
		(1.8, 1.2)
		-- (1.8, 2.2)
		-- (10.0, 2.2)
		-- (10.0, 1.2)
		-- (10.0, 2.2)
		-- (18.2, 2.2)
		-- (18.2, 1.2);
	\path[]	(1.8, 2.2)	-- (18.2, 2.2)	node[midway, above] () {Ab-square};
\end{tikzpicture}
\end{scaletikzpicturetowidth}
    \caption{The global structure of a fragment containing a well-placed Ab-square; there are three types of blocks: $c\BB,\, c\SS^I_i,\,\AA c$, where 
    $c$ is one of $\xx,\yy$. The blocks of the second type (which can be considered as essential blocks) are in color, each block is of length $M+1$ (recall that $|\AA|=|\BB|=|\SS_i|=M$). 
    The  special letters $\xx,\yy$ force each half of a well-placed Ab-square to contain an odd number of full 
    $\SS_i$'s and does not contain any $\SS_i$ only partially. 
    }\label{fig:prow}
\end{figure}

\begin{theorem}\label{thm:centers}
Computing all positions that are centers of Ab-square factors in a length-$n$ string over an alphabet of size $\omega(1)$ is $\ThreeSUM$-hard.
\end{theorem}
\begin{proof}
Due to \cref{Conv->Mid} we can reduce $\ConvThreeSUM$ in $N^{1+o(1)}$ time to an odd-half instance $\bar{x}$ of $\MidSUMthree$ of size $n=N^{1+o(1)}$ with elements in the range $[0,N^{3+o(1)}]$.

We construct the string $T$ as shown in Eq.~\ref{eqT} for the sequence $\bar{x}$. Then \cref{lem:well-placed} implies that $\MidSUMthree$ is a YES-instance if and only if $T$ has a well-placed Ab-square.
The string $T$ has length $\Oh(N^{1+o(1)}M)$. Each of the strings $\AA,\BB$ has length $M$ and is composed of $m=4^k$ strings of length $\Oh(kd)$, i.e., $\Psi$-images of $\Gamma$-strings. 

Hence, $M = \Oh(4^kkd)$. We select $k$ such that $k=\omega(1)$ and simultaneously $k=\Oh(\log N/\log \log N)$. Then we have $4^kk=N^{o(1)}$ and the $k$ primes are of magnitude $d=\Oh(N^{(3+o(1))/k})=N^{o(1)}$ (we can choose $k$ consecutive primes computed using Eratosthenes's sieve). 

Overall $|T|=N^{1+o(1)}$ and $|\Alphabet(T)|\le k+5 = \omega(1)$. (One can obtain any alphabet up to $\Oh(N)$ by appending distinct letters to $T$.)
\end{proof}

With the same argument for a constant-sized alphabet we obtain the following result.

\begin{theorem}\label{cor:centers}
All positions that are centers of Ab-square factors in a length-$n$ string over an alphabet of size $5+k$, for a constant $k$, cannot be computed in $\Oh(n^{2-\tfrac{6}{3+k}-\varepsilon})$ time, for a constant $\varepsilon>0$, unless the $\ThreeSUM$ conjecture fails.
\end{theorem}

\section{Computing centers of Ab-squares for constant-sized alphabets}
A set of vectors in $[1,n]^d$ is called \emph{monotone} if its elements can be ordered so that they form a monotone non-decreasing sequence on each coordinate.

\begin{definition}\label{def:AplusB}
For sets $\mathcal{A}$ and $\mathcal{B}$ of vectors we define
$$\mathcal{A}+ \mathcal{B}=\{a+b\,:\, a\in \mathcal{A}, b\in \mathcal{B}\},\ c\cdot \mathcal{A} = \{ca\,:\,a \in \mathcal{A}\}$$
and for a string $W$ we define:
$P_{l,r}(W) = \{ \Parikh(W[1..k]): l \le k \le r \}$.
Let us also denote by $|A|$ the length of a string corresponding to a Parikh vector $A$.
\end{definition}

In the algorithm we use the following fact shown in \cite{DBLP:conf/stoc/ChanL15}. 
The exact complexities can be found in \cite[Theorem 3.1]{DBLP:conf/stoc/ChanL15}.

\begin{fact}[\cite{DBLP:conf/stoc/ChanL15}]
\label{cor:stoc2015}
Given three monotone sequences $\mathcal{A},\mathcal{B},\mathcal{C}$ in $[1,n]^d$ for a constant $d$, we can compute $(\mathcal{A}+ \mathcal{B})
\cap \mathcal{C}$ in $\Oh(n^{2-\epsilon})$ expected time for a constant $\epsilon>0$, or in $\Oh(n^{2-\epsilon'})$ worst case time for a constant $\epsilon'>0$ if $d \le 7$.
\end{fact}

\newcommand{\Centers}{\mathsf{CENTERS}}
\newcommand{\lleft}{\mathit{left}}
\newcommand{\rright}{\mathit{right}}

\begin{algorithm}
\caption{$\Centers(T)$}\label{algo}
\textbf{if} {$|T|<2$} \textbf{return} $\emptyset$\; \vspace*{0.2cm}
$m=\lceil n/2 \rceil$\; \vspace*{0.2cm}
$\mathcal{A}=P_{0,m-1}(T)$;\ $\mathcal{B}=P_{m,n}(T)$;\ $\mathcal{C}=  P_{0,n}(T)$\;
\vspace*{0.2cm}
$\mathcal{M} = \{|C|\,:\, 2C\in (\mathcal{A}+ \mathcal{B})\cap 2\cdot \mathcal{C}\}$\;
\vspace*{0.2cm}
$T_{\lleft}=T[1..m-1]$;\ $T_{\rright}=T[m..n]$\;\vspace*{0.2cm}
\KwRet{$\mathcal{M} \cup \Centers(T_{\lleft})\cup \{k+m\,:\, k\in\Centers(T_{\rright})\}$}

\end{algorithm}

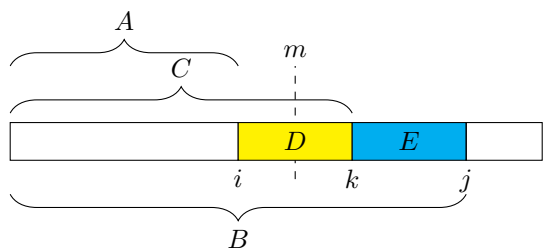
\begin{figure}[h!!]
\centering
\begin{tikzpicture}[scale=0.5]
	\path[draw, dashed]	(7.5, -0.5)	-- (7.5, 2.5)	node[above] () {$m$};
	\path[draw]	(0.0, 0.0)	rectangle (6.0, 1.0);
	\path[draw, fill=yellow]	(6.0, 0.0)	rectangle (9.0, 1.0)	node[pos=.5, color=black] () {$D$};
	\path[draw, fill=cyan]	(9.0, 0.0)	rectangle (12.0, 1.0)	node[pos=.5, color=black] () {$E$};
	\path[draw]	(12.0, 0.0)	rectangle (14.0, 1.0);
	\path[]
		(6.0, 0.0)	node[below] () {$i$}
		(9.0, 0.0)	node[below] () {$k$}
		(12.0, 0.0)	node[below] () {$j$};
	\path[draw, decorate,decoration={brace, amplitude=10pt}]	(0.0, 2.5)	-- (6.0, 2.5)	node[above=10pt, midway] () {$A$};
	\path[draw, decorate,decoration={brace, amplitude=10pt}]	(0.0, 1.25)	-- (9.0, 1.25)	node[above=10pt, midway] () {$C$};
	\path[draw, decorate,decoration={brace, amplitude=10pt, mirror}]	(0.0, -0.9)	-- (12.0, -0.9)	node[below=10pt , midway] () {$B$};
\end{tikzpicture}
\caption{$A\in \mathcal{A}$, $B\in \mathcal{B}$, $C\in \mathcal{C}$ denote Parikh vectors of the corresponding fragments. If $A+B=2C$, then $D=E$ and $k=|C|$ is a center of an Ab-square.
}
\label{fig:parikh-vectors-of-abelian-square}
\end{figure}

\begin{theorem}\label{thm:up}
For a string of length $n$ over an alphabet of size $d=\Oh(1)$, we can compute centers of all 
Ab-squares  and centers of all odd Ab-squares in expected time $\Oh(n^{2-\epsilon})$ or 
in worst case time $\Oh(n^{2-\epsilon})$ if $d \le 7$, for $\epsilon >0$.
\end{theorem}
\begin{proof}
We use the above algorithm.
Correctness of the algorithm is straightforward; see Figure~\ref{fig:parikh-vectors-of-abelian-square}.
 If $$|A|<|B|, \ |C|=(|A|+|B|)/2, \ B=A+D+E,\ C=A+D$$ then $A+B=2C\iff  2A+D+E=2A+2D$. 
 
 Consequently, after
cancelling the same parts on both sides, $A+B=2C\iff E=D$, equivalently if and only if the factor $T[i..j]$ corresponding to $DE$ is an Ab-square centred in $k=|C|$.
The figure shows the case when $k$ is in the right half of the strings; the other case is symmetric.

By \cref{cor:stoc2015} the cost of the algorithm can be given by a recurrence \[S(n)=2\cdot S\left(\tfrac{n}{2}\right)+\Oh(n^{2-\epsilon})\] which results in $S(n)=\Oh(n^{2-\epsilon})$ for  $\epsilon>0$.

In case of of odd Ab-squares let
 \[P^c_{l,r}(W) = \{ \Parikh(W[1..k]): l \le k \le r,\,k \bmod{2} = c \}.\]  In the algorithm the statement $\mathcal{M} = \{|C|\,:\, 2C\in (\mathcal{A}+ \mathcal{B})\cap 2\cdot \mathcal{C}\}$
 is executed for both $c \in \{0,1\}$, with
\[\mathcal{A}=P^c_{0,m-1}(T),\quad \mathcal{B}=P^c_{m,n}(T),\quad \mathcal{C}= P^{1-c}_{0,n}(T).\]
Other parts of the algorithm, as well as its analysis, are essentially the same.
\end{proof}

\section{Detecting odd Ab-squares}
Unfortunately the string $T$ from \cref{lem:well-placed} has many Ab-squares which are
not well-placed. Our approach is to embed the (slightly) modified string 
$T$ into a string which is a special composition  of $T$ and 
a combination of long quaternary Ab-square-free strings. The resulting string will fix the potential centers in specified locations.
We use additional letters: $\diamondsuit,\, \circ$ and $\zero,\dots,\six$.

\subsection{Fixing centers}
We show first a fact useful in fixing Ab-squares in specified places (Lemma~\ref{lem:zip}). Ker\"anen's construction~\cite{DBLP:conf/icalp/Keranen92} of a quaternary Ab-square-free string consists in iterating a certain morphism $\phi$, such that $|\phi(a)|=85$ for each of the four letters $a$, on an initially single-letter string. This implies the following lemma.
\begin{lemma}[Ker\"anen~\cite{DBLP:conf/icalp/Keranen92}]\label{lem:keranen}
A length-$n$ quaternary Ab-square-free string can be generated in $\Oh(n)$ time.
\end{lemma}

Let $P_{t-2}$ be any Ab-square-free string of length $t-2$ over alphabet $\{\three,\four,\five,\six\}$.
 Let us define
\[U_{2t}\,=\,\zero\,P_{t-2}\,\one\,\two\,P_{t-2}^R\,\zero.\]

\begin{figure}[htpb]
\begin{center}
    \begin{tikzpicture}[scale=0.43]
        \draw[thick] (0,0) -- (5,5) -- (10,0) -- (15,5) -- (20,0) -- (25,5);
        \draw[line width=0.1cm,blue] (4,4) -- node[above left=-0.2cm] {\textcolor{black}{$A$}} (5,5);
        \begin{scope}[xshift=10cm]
            \draw[line width=0.1cm,blue] (4,4) -- node[above left=-0.2cm] {\textcolor{black}{$A$}} (5,5);
            \draw[line width=0.1cm,red] (0,0) -- node[above left=-0.2cm] {\textcolor{black}{$B$}} (2,2);
            \draw[line width=0.1cm,black] (2,2) -- node[above left=-0.2cm] {\textcolor{black}{$C$}} (3,3);
            \draw[line width=0.1cm,olive] (3,3) -- node[above left=-0.2cm] {\textcolor{black}{$D$}} (4,4);
        \end{scope}
        \draw[xshift=20cm,line width=0.1cm,red] (0,0) -- node[above left=-0.2cm] {\textcolor{black}{$B$}} (2,2);
        \draw[densely dotted] (2,2) -- (22,2) (3,3) -- (23,3) (4,4) -- (24,4);
        \draw (4,4) -- (4,-1.5) -- node[below] {supposed Ab-square} (22,-1.5) -- (22,2)  (13,3) -- (13,-1.5);
        \draw[snake=brace,xshift=-0.7cm,yshift=0.7cm] (0,0) -- node[above left=-0.05cm] {$P$} (5,5);
        \draw[snake=brace,xshift=0.1cm,yshift=0.1cm] (15,5) -- node[above right=-0.05cm] {$P^R$} (20,0);
        \draw (5,5) node[above] {\footnotesize $\one\,\two$};
        \draw (15,5) node[above] {\footnotesize $\one\,\two$};
        \draw (10,0) node[below] {\footnotesize $\zero\,\zero$};
        \draw (20,0) node[below] {\footnotesize $\zero\,\zero$};
    \end{tikzpicture}
\end{center}
\centering
\caption{Illustration of the proof that  Ab-square cannot start inside $P=P_{t-2}$
and end inside another instance of $P$, not having its center between two $\zero$'s and have length which is not an even multiple of the period $2t$.
If it is an Ab-square (as shown in the figure) then $A\one\two P^R\zero\zero BC\cong DA\one\two P^R\zero\zero B$,
%which implies $CA\one\two P^R\zero\zero B\cong DA\one\two P^R\zero\zero B$ %(since Abelian concatenation is commutative),
then we can cancel equal parts on both sides.
Consequently $ C\cong D$, which implies that $ P^R$ contains an Ab-square $CD$; a contradiction.}\label{fig:gory}
\end{figure}
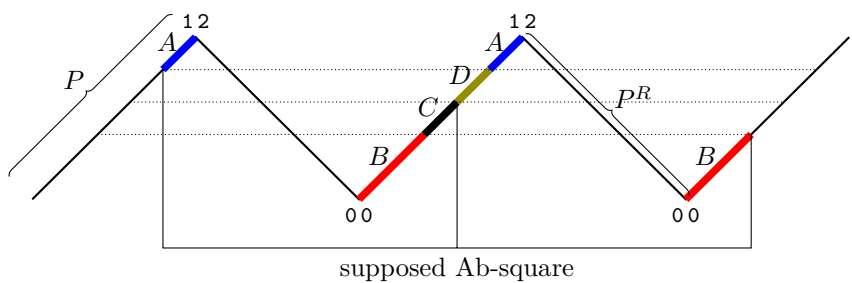

\begin{lemma}\label{lem:usztyw}
The string $(U_{2t})^m$ contains exactly the following Ab-squares: 
\begin{enumerate}[(1)]
\item of length divisible by $4t$; and 
\item with the center between two $\zero$'s and of all admissible even lengths other than $(4q+2)t$, 
 for an integer $q \ge 0$.
\end{enumerate}
\end{lemma}
\begin{proof}
Let $X$ be a factor of $(U_{2t})^m$. If $X$ has length greater than or equal to $4t$, then its middle length-$4t$ factor forms a classic square, and after removing it we obtain a different factor $Y$ of length smaller by $4t$, which is centred exactly like $X$. Hence, we can focus only on non-empty factors of length smaller than $4t$.

If $X$ is centred between two $\zero$'s, then after removing letters $\one$ and $\two$ we obtain an even palindrome (hence also an Ab-square). If $X$ is shorter than $2t$, then no letters $\one$ or $\two$ occur. If it is longer than $2t$, then both parts contain one letter $\one$ and $\two$ each. If its length is exactly $2t$, then the letters $\one$ and $\two$ remain unmatched, hence it is the only case where the factor is not an Ab-square.

Let us assume that $X$ is centred in a different place. If the factor does not contain any of the letters $\zero$, $\one$ or $\two$, then it is a factor of $P_{t-2}$ or its reverse, hence it cannot be an Ab-square. Otherwise, the factor needs to contain each of $\one$, $\two$ twice and $\zero$ four times. Then $X$ fully contains a factor

\smallskip\centerline{
$\one\,\two\, P_{t-2}^R\, \zero\,\zero\, P_{t-2}\, \one\,\two\, P_{t-2}^R\, \zero\,\zero\quad\text{or}\quad\zero\,\zero\, P_{t-2}\, \one\,\two\, \,P_{t-2}^R\, \zero\,\zero\, P_{t-2}\, \one\,\two.$}

\smallskip
Let us assume the former, see Figure~\ref{fig:gory}; the latter is considered analogously. String $X$ has a length-$i$ suffix of $P_{t-2}$ as a prefix and a length-$j$ prefix of $P_{t-2}$ as a suffix, with $0 \le i+j \le t-2$ and $t-2-(i+j)$ even. Then $X$ is an Ab-square if and only if $P_{t-2}[j+1..t-2-i]$ is an Ab-square, since $P_{t-2}[(t-i+j)/2..t-2-i]\,X\, P[j+1..(t-i+j)/2-1]$ is a classic square; see \cref{fig:gory}.
\end{proof}

\begin{remark}
\cref{lem:usztyw} works for any Ab-square-free string $P_{t-2}$ such that $\Alphabet(P_{t-2}) \cap \{\zero,\one,\two\}=\emptyset$.
\end{remark}

For equal-length strings $X,Y$ we define the string 
\newcommand{\shuffle}{\mathsf{shuffle}}
\[
\shuffle_\zz(X,Y)\,=\, X[1]\,\zz\,Y[1]\;X[2]\,\zz\,Y[2]\;X[3]\,\zz\,Y[3]\; \cdots .\]
For example, $\shuffle_\diamondsuit(\mathtt{abc},\mathtt{ABC})\,=\,\mathtt{a}\diamondsuit\mathtt{Ab}\diamondsuit\mathtt{Bc}\diamondsuit\mathtt{C}$.

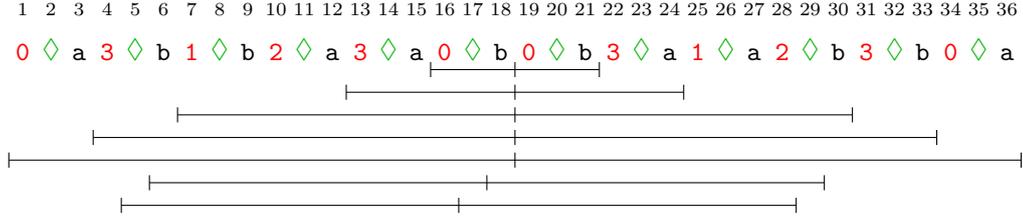
\begin{figure}[t]
    \centering
\begin{tikzpicture}[xscale=0.37]
    \foreach \x in {1,...,36}{
        \draw (\x,0.6) node[above] {\scriptsize{\x}};
    }
    \foreach \dx in {0,18}{
        \foreach \x/\c in {
            1/0,4/3,7/1,10/2,13/3,16/0,
        }{
            \draw[xshift=\dx cm] (\x,0) node[above] {\textcolor{red}{\texttt{\c}}};
        }
        \foreach \x in {
            2,5,8,11,14,17
        }{
            \draw[xshift=\dx cm] (\x,0) node[above] {\textcolor{green!70!black}{$\zz$}};
        }
    }
    \foreach \x/\c in {
        3/a,6/b,9/b,12/a,15/a,18/b,
        21/b,24/a,27/a,30/b,33/b,36/a
    }{
        \draw (\x,0) node[above] {\texttt{\c}};
    }
    \foreach \l/\m/\r/\h in {
        15.5/18.5/21.5/0,
        12.5/18.5/24.5/-0.3,
        6.5/18.5/30.5/-0.6,
        3.5/18.5/33.5/-0.9,
        0.5/18.5/36.5/-1.2,
        5.5/17.5/29.5/-1.5,
        4.5/16.5/28.5/-1.8
    }{
        \draw (\l,\h-0.1) -- (\l,\h+0.1);
        \draw (\m,\h-0.1) -- (\m,\h+0.1);
        \draw (\r,\h-0.1) -- (\r,\h+0.1);
        \draw (\l,\h) -- (\r,\h);
    }
\end{tikzpicture}
    \caption{Illustration of \cref{lem:zip}. 
    Let $X=(U_6)^2$. The string $Y=(\mathtt{abba})^3$, composed of black letters, contains many Ab-squares. However the string $Z=\shuffle_\zz(X,Y)$ of length 36, shown above,  contains only Ab-squares centred at 16, 17 or 18, as in the figure.
    The implied Ab-squares in $Z$ are only those which are 
    centred at positions 5 or 6 in $Z$.}
    \label{fig:zip}
\end{figure}

The parity condition for half lengths of Ab-squares in the following observation justifies the usage of the additional letter $\zz$ in $\shuffle$.
Let $U_{[X]}$ be the string resulting from $U$ by removing all letters outside $\Alphabet(X)$.
\begin{observation}\label{obs:filtr}
Assume $X,Y$ are equal-length strings composed of disjoint sets of letters distinct from $\zz$ and $W$ is an Ab-square in $\shuffle_\zz(X,Y)$.
Then $W_{[X]},W_{[Y]}$ are Ab-squares in $X,Y$, respectively (we say that 
these Ab-squares are \emph{implied} by $W$). Moreover, $|W_{[X]}|/2,\, |W_{[Y]}|/2,\, W|/2$ are of the same parity.
\end{observation}
We say that an even-length factor of a string $X$ is \emph{centred at} $i$ if it has its center between positions $i$ and $i+1$ in $X$.
By $a \mid b$ and $a \nmid b$ we denote that $a$ divides $b$ and $a$ does not divide~$b$.
For an illustration of the following lemma, see \cref{fig:zip}.

\begin{lemma}\label{lem:zip}
Let $X=(U_{2t})^{n-1}$, $Y$ be a string of length $|X|$ such that its alphabet
is disjoint with $\Alphabet(X) \cup \{\zz\}$, $W=\shuffle_\zz(X,Y)$, and let an integer $\ell$ satisfy $12t \nmid \ell$.
Then a length-$\ell$ factor of  $W$ is an Ab-square if and only if it is centred in $W$ at $r \equiv \{0,-1,-2\} \pmod{6t}$, $Y$ contains an Ab-square factor 
of length $\ell/3$ centred in $Y$ at $\floor{r/3}$, and $6t \nmid \ell$.
\end{lemma}

\begin{proof}
By the disjointness of sets of letters in $X,Y$ and $\{\zz\}$, each Ab-square in $W$ has length that is divisible by 3. The following claim is then readily verified (cf.\ \cref{obs:filtr}).
\begin{claim}
For positive integer $\ell$ such that $6 \mid \ell$, a length-$\ell$ factor of $W$ centred at $r$ is an Ab-square if and only if the length-$\ell/3$ factors in $X$ and $Y$ centred at $\floor{\tfrac{r+2}{3}}$ and $\floor{\tfrac{r}{3}}$, respectively, are Ab-squares.
\end{claim}
Let integer $\ell>0$ satisfy $6 \mid \ell$ and $12t \nmid \ell$.
We show two implications.

\medskip\noindent
$(\mathbf{\Rightarrow})$
If $W$ contains an Ab-square factor of length $\ell$ centred at some $r$, then the implied Ab-square factor of $X$ has length $\ell/3$, where $4t \nmid \ell/3$, so by \cref{lem:usztyw} it has its center between two $\zero$'s, i.e., $2t \mid \floor{\tfrac{r+2}{3}}$. Hence, $r \equiv \{0,-1,-2\} \pmod{6t}$. 

Moreover, $2t \nmid \ell/3$ also by \cref{lem:usztyw}. Finally, the implied Ab-square factor of $Y$ indeed has length $\ell/3$ and is centred at $\floor{r/3}$.

\medskip\noindent
$(\mathbf{\Leftarrow})$
Let $r \equiv \{0,-1,-2\} \pmod{6t}$, $6t \nmid \ell$, and assume that $Y$ contains an Ab-square factor of length $\ell/3$ centred at $\floor{r/3}$.
We have $2t \mid \floor{\tfrac{r+2}{3}}\ \text{and} \ 2t \nmid \ell/3,$
so by \cref{lem:usztyw} the string $X$ contains an Ab-square factor of length $\ell/3$ centred at $\floor{\tfrac{r+2}{3}}$. 
Finally, the unary string $\zz^{2tm}$, certainly contains an Ab-square factor of length $\ell/3$ centred at $\floor{\tfrac{r+1}{3}}$. By the claim, $W$ contains an Ab-square of length $\ell$ centred at $r$ that implies the three Ab-squares.
\end{proof}

\subsection{Main result}
We use the technique of fixing Ab-squares from \cref{lem:zip}. Moreover, we make the following minor modifications upon the construction of 
string $T$ in \cref{subsec:centers}: 
\begin{enumerate}[(1)]
\item\label{ito} Each fragment $\bb^{kd}$ is extended by one letter to $\bb^{kd+1}$, and 

\item\label{itt} the letters $\xx,\yy$ are replaced each by two letters $\xx\zez,\yy\zez$, respectively. 
\end{enumerate}
Intuitively, \eqref{ito} allows to extend Ab-squares considered in the proof of \cref{lem:well-placed} by one letter $\bb$ to either side, and \eqref{itt} makes $|\AA|=|\BB|=|\SS_i^I|$ even which facilitates the usage of \cref{lem:zip} with $Y=T$.
It can be verified by inspecting the proof that \cref{lem:well-placed} still holds after these two changes. We refer to all the notions from \cref{subsec:centers} after these modifications.

\begin{theorem}\label{thm:lengths}
Checking if a length-$n$ string over an alphabet of size $\omega(1)$ contains an odd Ab-square is $\ThreeSUM$-hard. Moreover, for a string over an alphabet of size $14+k$, for a constant $k$, the same problem cannot be solved in $\Oh(n^{2-\tfrac{6}{3+k}-\varepsilon})$ time, for a constant $\varepsilon>0$, unless the $\ThreeSUM$ conjecture fails.
\end{theorem}
\begin{proof}
It is enough now to show the following equivalence  for $X=(U_{2t})^{n-1}$, where $2t=|T|/(n-1)$. We assume that $n \ge 3$.
\begin{claim}\label{clm:lengths}
 An odd-half instance of $\MidSUMthree$ is a YES-instance if and only if $W=\shuffle_\zz(X,T)$ has an odd Ab-square  factor.
\end{claim}
\begin{proof} 

\begin{figure}[h!]
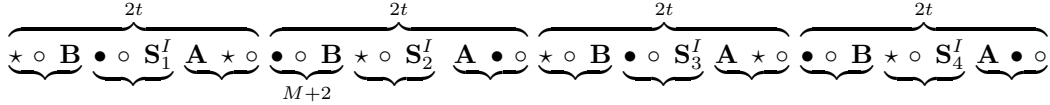

\centering
$
    \overbrace{\underbrace{\yy\,\zez\;\BB}\; \underbrace{\xx\,\zez\;\SS^I_1}\; \underbrace{\AA\;\yy\,\zez}}^{2t}\;
   \overbrace{ \underbrace{\xx\,\zez\;\BB}_{M+2}\;  \underbrace{\yy\,\zez\;\SS^I_2}\; \;\underbrace{\AA\;\xx\,\zez}}^{2t}\;
    \overbrace{\underbrace{\yy\,\zez\;\BB}\; \underbrace{\xx\,\zez\;\SS^I_3}\; 
    \underbrace{\AA\;\yy\,\zez}}^{2t}\;
    \overbrace{\underbrace{\xx\,\zez\;\BB}\; \underbrace{\yy\,\zez\;\SS^I_4}\; 
    \underbrace{\AA\,\xx\,\zez}}^{2t}
$
\caption{A schematic structure of a fragment of $T$ after insertion of symbols $\zez$.
 There are $3(n-1)$ (underbraced) blocks in $T$, each  of size $M+2$, and $2t=3M+6$.}\label{dawnefig5}
\end{figure}

%We show two implications.

%\smallskip\noindent
$(\mathbf{\Rightarrow})$ 
Assume that $\bar{x}$ is an odd-half instance and $\MidSUMthree(\bar{x})$ has a solution. 

By \cref{lem:well-placed}, $T$ contains a well-placed Ab-square, that is, an Ab-square centred at a position $r'$ such that $2t \mid r'$. (Recall that $3(M+2)=2t$.) Moreover, in the proof of that lemma it is shown that in this case there exists a well-placed Ab-square in $T$ that satisfies the following additional requirements: 
\textbf{(1)} it starts within the gadget $\BB$;\
\textbf{(2)} it starts and ends within a block of $\bb$'s; 
\ \textbf{(3)} its maximal prefix and suffix consisting of letters $\bb$ are $\bb^e$ and $\bb^f$, where $e,f \le kd$. 

Let $\ell'$ denote the half length of this Ab-square. By \textbf{(2)} and \textbf{(3)}, if $\ell'$ is even, the Ab-square can be extended by one letter $\bb$ to either side (because we have extended each block $\bb^{kd}$) so that $\ell'$ becomes odd. Moreover, by \textbf{(1)}, we have $\ell' \bmod (2t) \in [\tfrac43t,2t)$, in particular, $t\nmid\ell'$. 
Then \cref{lem:zip} concludes that the factor of $W$ centred at $r=3r' \equiv 0 \pmod{6t}$ and of length $6\ell'$ such that $6t \nmid 6\ell'$ is an Ab-square. Its half length, $3\ell'$, is odd, as desired.

\begin{figure}[h!]
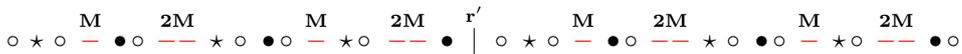

    \centering
$\zez\,\yy\,\zez\,\stackrel{\mathbf{M}}{\textcolor{red}{\mathbf -}}\, \xx\,\zez\,\stackrel{\mathbf{2M}}{\textcolor{red}{\mathbf -}\textcolor{red}{\mathbf -}}\,\yy\,\zez\,\,
    \xx\,\zez\,\stackrel{\mathbf{M}}{\textcolor{red}{\mathbf -}}\,  \yy\,\zez\,\,\stackrel{\mathbf{2M}}{\textcolor{red}{\mathbf -}\textcolor{red}{\mathbf -}}\,\xx\,
    \stackrel{\mathbf{r'}}{\vert}\,\zez\,\,
    \yy\,\zez\,\stackrel{\mathbf{M}}{\textcolor{red}{\mathbf -}}\, \xx\,\zez\,\stackrel{\mathbf{2M}}{\textcolor{red}{\mathbf -}\textcolor{red}{\mathbf -}}\,
    \yy\,\zez\,\,
    \xx\,\zez\,\stackrel{\mathbf{M}}{\textcolor{red}{\mathbf -}}\, \yy\,\zez\,\stackrel{\mathbf{2M}}{\textcolor{red}{\mathbf -}\textcolor{red}{\mathbf -}}
    \,\xx\,\zez$
    \caption{A simplified version of \cref{dawnefig5}. 
    Which position in a block can be the starting position of an Ab-square with the  center at $r'$ (one position to the left of a \emph{good} center), only counting  $\xx,\yy,\zez$.
    }
    \label{fig:xyz}
\end{figure}

\medskip\noindent
$(\mathbf{\Leftarrow})$ 
Assume that $W$ has an Ab-square factor $U$ of length $\ell$ such that $\ell/2$ is odd. In particular, we have $12t \nmid \ell$, so by \cref{lem:zip} the Ab-square $U$ is 
centred in $W$ at $r \equiv \{0,-1,-2\} \pmod{6t}$ and $T$ contains an Ab-square factor $V$ of length $\ell/3$ centred in $T$ at $r'=\floor{r/3}$. If $6t \mid r$, then $2t \mid r'$ and $V$ is well-placed. 

Otherwise, $V$ cannot be an Ab-square due to the following fact:
$T$ does not contain an Ab-square factor of length $\ell$ not divisible by $4t$ and centred at $r' \equiv -1 \pmod{2t}$.
Indeed, similarly as in the proof of \cref{lem:well-placed}, we will show that each even-length factor centred at such $r'$ contains different counts of one of the letters $\xx,\yy,\zez$ in both halves.
The positions of letters $\xx,\yy,\zez$ in $T$ repeat with period $6(M+2)$, so it is sufficient to inspect the first 6 blocks on each side, as the remaining ones will behave periodically; see \cref{dawnefig5,fig:xyz}.

\begin{table}[htpb]
    \centering
    \vspace*{0.3cm}
    \begin{tabular}{r*{9}{|c}}
      letter & $\zez$ & $\yy$ & $\zez$ & $\BB$ & $\xx$ & $\zez$ & $\SS$ & $\AA$ & $\yy$  \\
      distance & $6M+12$ & $6M+11$ & $6M+10$ & & $5M+9$ & $5M+8$ & & & $3M+7$ \\
      &&&&&&&&&\\
      letter & $\zez$ & $\xx$ & $\zez$ & $\BB$ & $\yy$ & $\zez$ & $\SS$ & $\AA$ & $\xx$ \\
      distance & $3M+6$ & $3M+5$ & $3M+4$ & & $2M+3$ & $2M+2$ & & & $1$
    \end{tabular}
    
    \bigskip
    \begin{tabular}{r*{9}{|c}}
      letter & $\zez$ & $\yy$ & $\zez$ & $\BB$ & $\xx$ & $\zez$ & $\SS$ & $\AA$ & $\yy$  \\
      distance & $1$ & $2$ & $3$ & & $M+4$ & $M+5$ & & & $3M+6$ \\
      &&&&&&&&&\\
      letter & $\zez$ & $\xx$ & $\zez$ & $\BB$ & $\yy$ & $\zez$ & $\SS$ & $\AA$ & $\xx$ \\
      distance & $3M+7$ & $3M+8$ & $3M+9$ & & $4M+10$ & $4M+11$ & & & $6M+12$
    \end{tabular}
    \vspace*{0.4cm}
    \caption{Top/bottom table: the distances of letters from $\{\xx,\yy,\zez\}$ from the left/right half to the center of the factor.}
    \label{tab:distances}
\end{table}

An exhaustive verification can be performed as follows. First, in \cref{tab:distances}, we count the distances of letters from $\{\xx,\yy,\zez\}$ in both directions to the center of the factor. In \cref{tab:xyz} we perform a merge of these two sequences of distances assuming that $M \ge 3$. 

For each distance, we write a letter that is located at this distance with a ``$+$'' sign if it is in the left half and with a ``$-$'' sign otherwise. Then remaining columns show the partial sum of the number of occurrences of the letter $c$ in the left and in the right half, for each $c \in \{\xx,\yy,\zez\}$.

\begin{table}[htpb]
    \centering
    \twocol{
    \begin{tabular}{c|c|c|c|c}
      position & letter & $\xx$ & $\yy$ & $\zez$ \\\hline
      $1$ & $+\xx$ &$1$&$0$&$0$\\
      $1$ & $-\zez$ &$1$&$0$&$-1$\\
      $2$ & $-\yy$ &$1$&$-1$&$-1$\\
      $3$ & $-\zez$ &$1$&$-1$&$-2$\\
      $M+4$ & $-\xx$ &$0$&$-1$&$-2$\\
      $M+5$ & $-\zez$ &$0$&$-1$&$-3$\\
      $2M+2$ & $+\zez$ &$0$&$-1$&$-2$\\
      $2M+3$ & $+\yy$ &$0$&$0$&$-2$\\
      $3M+4$ & $+\zez$ &$0$&$0$&$-1$\\
      $3M+5$ & $+\xx$ &$1$&$0$&$-1$\\
      $3M+6$ & $+\zez$ &$1$&$0$&$0$\\
      $3M+6$ & $-\yy$ &$1$&$-1$&$0$\\
    \end{tabular}
    }{
    \begin{tabular}{c|c|c|c|c}
      position & letter & $\xx$ & $\yy$ & $\zez$ \\\hline
      $3M+7$ & $+\yy$ &$1$&$0$&$0$\\
      $3M+7$ & $-\zez$ &$1$&$0$&$-1$\\
      $3M+8$ & $-\xx$ &$0$&$0$&$-1$\\
      $3M+9$ & $-\zez$ &$0$&$0$&$-2$\\
      $4M+10$ & $-\yy$ &$0$&$-1$&$-2$\\
      $4M+11$ & $-\zez$ &$0$&$-1$&$-3$\\
      $5M+8$ & $+\zez$ &$0$&$-1$&$-2$\\
      $5M+9$ & $+\xx$ &$1$&$-1$&$-2$\\
      $6M+10$ & $+\zez$ &$1$&$-1$&$-1$\\
      $6M+11$ & $+\yy$ &$1$&$0$&$-1$\\
      $6M+12$ & $+\zez$ &$1$&$0$&$0$\\
      $6M+12$ & $-\xx$ &$0$&$0$&$0$\\
    \end{tabular}
    }
    \vspace*{0.2cm}
    \caption{The merge of distance sequences from \cref{tab:distances}.
    }
    \label{tab:xyz}
\end{table}
%An exhaustive verification of several cases can be performed by counting  distances of letters from $\{\xx,\yy,\zez\}$ in both directions to the center of the factor (for more details see the full version).

Consequently, 
as in \cref{lem:well-placed}, the corresponding instance of $\MidSUMthree$ is a YES-instance.
\end{proof}
The complexities in the theorem are obtained as in \cref{thm:centers,cor:centers}.
\end{proof}

\section{Open problems}
The most interesting questions that remain open are as follows:

\begin{enumerate}
   \item Is checking Ab-square-freeness $\ThreeSUM$-hard? Our reductions allowed us to show $\ThreeSUM$-hardness of detecting an odd Ab-square.
   \item  Can one detect an additive square in a length-$n$ string over a constant-sized alphabet in $\Oh(n^{2-\varepsilon})$ time, for some $\varepsilon>0$? We have shown $\ThreeSUM$-hardness of this problem for an alphabet that is polynomial in $n$.
   \end{enumerate}

\bibliographystyle{plainurl}

\bibliography{reference}

\end{document}